\documentclass[11pt]{article}
\usepackage[preprint]{acl}
\usepackage{times}
\usepackage[T1]{fontenc}
\usepackage{microtype}
\usepackage{amsmath,amssymb,booktabs,array,tabularx,tikz}
\usepackage{graphicx}
\usetikzlibrary{arrows.meta}
\hypersetup{hidelinks}

\title{VerifierBench: Benchmarking the Generation of Executable Verifiers for LLM Agents}
\author{
  \bfseries Jianing Yin$^{1}$ \quad Ziyi Chen$^{2}$ \quad Hao Wang$^{3}$ \quad Yiyou Sun$^{3}$ \\[2pt]
  \bfseries Zijian Jin$^{4}$ \quad Dawn Song$^{3}$ \quad Daoan Zhang$^{5}$ \\[6pt]
  \normalfont
  $^{1}$University of Pennsylvania \quad
  $^{2}$University of Illinois at Urbana-Champaign \\
  $^{3}$University of California, Berkeley \quad
  $^{4}$New York University \quad
  $^{5}$University of Rochester
}

\begin{document}
\maketitle
\raggedbottom

\begin{abstract}
Agents need reliable ways to validate their outputs, both when completing tasks and when filtering synthetic training data.
We study verifier generation: from the instruction and environment available for solving a task, an agent writes an executable verifier that must accept every valid solution and reject invalid ones.
We introduce \textbf{VerifierBench}, a benchmark of 222 tasks from six agent benchmarks spanning tool-call, state-based, and code verification, with 4,993 hidden test cases that include alternative valid solutions and attacks; Verifier4Verifier (V4V), a six-check audit, screens its source verifiers and test suites.
Across 19 models in a common agent harness, normalized balanced accuracy ranges from 10.6 to 62.1, with generated verifiers consistently rejecting invalid solutions more reliably than they accept valid ones.
This apparent strength can mask overly restrictive verification: most false rejections in the analyzed records come from verifiers that accept no tested solution to a task.
Case studies reveal unsupported assumptions about solution formats, tool use, and code changes that exclude valid solutions.
These errors survive self-testing because agents test solutions that share the verifier's mistaken assumptions, or submit verifiers without confirming that any correct solution passes.
These findings suggest testing both whether a verifier accepts a correct solution and whether it accepts other solutions permitted by the task.
The project page%
\footnote{\raggedright\urlstyle{same}%
\url{https://verifierbench.github.io/}}
and dataset%
\footnote{\raggedright\fontsize{8}{9}\selectfont%
\mbox{\href{https://huggingface.co/datasets/Catherine1213/VerifierBench}%
{https://huggingface.co/datasets/Catherine1213/VerifierBench}}}
are publicly available.
Code is also publicly available%
\footnote{\raggedright\urlstyle{same}%
\url{https://github.com/DwanZhang-AI/verifier-suite}}.
\end{abstract}

\section{Introduction}
\label{sec:intro}

In data-oriented recursive self-improvement (RSI) \citep{yin2025godel, min2026tracedance}, agents construct training data for their own subsequent learning and can jointly generate tasks and verification functions that provide reinforcement learning rewards \citep{zhou2026self}.
These verifiers judge candidate solutions, determining which receive rewards or enter the training data.
An incorrect verifier can therefore repeatedly reward a violation or discard a valid solution, making the reliability of agent-generated verifiers critical to this learning process.

To examine this reliability, we study \emph{verifier generation} \citep{zhang2025generative, dinella2022toga}: given the instruction and environment a solver would receive, an agent writes an executable verifier that returns \texttt{PASS} or \texttt{FAIL} for any candidate solution.
Unlike a solver, which needs one valid solution, a verifier must accept all valid solutions while rejecting invalid ones.
It must therefore distinguish task requirements from choices that can vary across valid solutions.
Figure~\ref{vb:rule-case} shows a verifier rejecting a permitted read-only query because the agent omits its tool from an allowlist.
The self-tests omit the same query and never establish that a complete valid solution passes.

A verifier that accepts one expected solution may still reject other valid ways to complete a task, while one that catches ordinary mistakes may still accept solutions that exploit gaps in its checks.
Evaluating verifier generation therefore requires testing both acceptance of valid alternatives and rejection of invalid solutions, including deliberate attacks.
Prior work evaluates generated test assertions \citep{dinella2022toga}, unit tests \citep{jain2025testgeneval}, issue-reproducing tests \citep{mundler2024swt}, and executable specifications \citep{endres2024can, chen2026codespecbench}.
We extend this line of evaluation to agent task completion, where correctness can depend on tool interactions, resulting environment states, and code changes.
To this end, we build \textbf{VerifierBench}: 222 tasks from six open-source agent benchmarks with 4,993 hidden test cases, comprising reference solutions, valid alternatives, ordinary errors, and attacks on verification.
This design tests whether generated verifiers distinguish task success from failure across different ways of carrying out the same task.
Reliable measurement also requires trustworthy evaluation labels.
Because both the benchmarks' source verifiers and our constructed test cases can contain errors, we audit them with \textbf{Verifier4Verifier (V4V)}, six checks that include replay, repeatability, and mutation testing.
Source verifiers that pass this audit serve as the reference verifiers.

Evaluating 19 models in a common harness and 17 further model--harness configurations, we report three findings.
First, generated verifiers accept valid solutions less reliably than they reject invalid ones: as reported in Section~\ref{sec:overall}, true positive rates range from 21.1\% to 70.2\%, whereas true negative rates range from 79.2\% to 96.3\%.
Second, as analyzed in Section~\ref{sec:errors}, these false rejections concentrate in \emph{all-reject verifiers}, which return \texttt{FAIL} on every test case of their task and produce 87.9--96.0\% of false rejections in the analyzed records.
Third, in selected cases examined in Section~\ref{sec:analysis}, these verifiers impose checks that the task does not require, and the agents' self-tests either exercise only solutions those checks allow or never confirm that any valid solution passes.
We thus contribute a benchmark and screening procedure for verifier generation in Section~\ref{sec:benchmark}, together with an analysis in Section~\ref{sec:implications} that motivates testing generated verifiers on permitted variations, not only on a chosen solution.

\begin{figure*}[t]
\centering
\includegraphics[width=\textwidth]{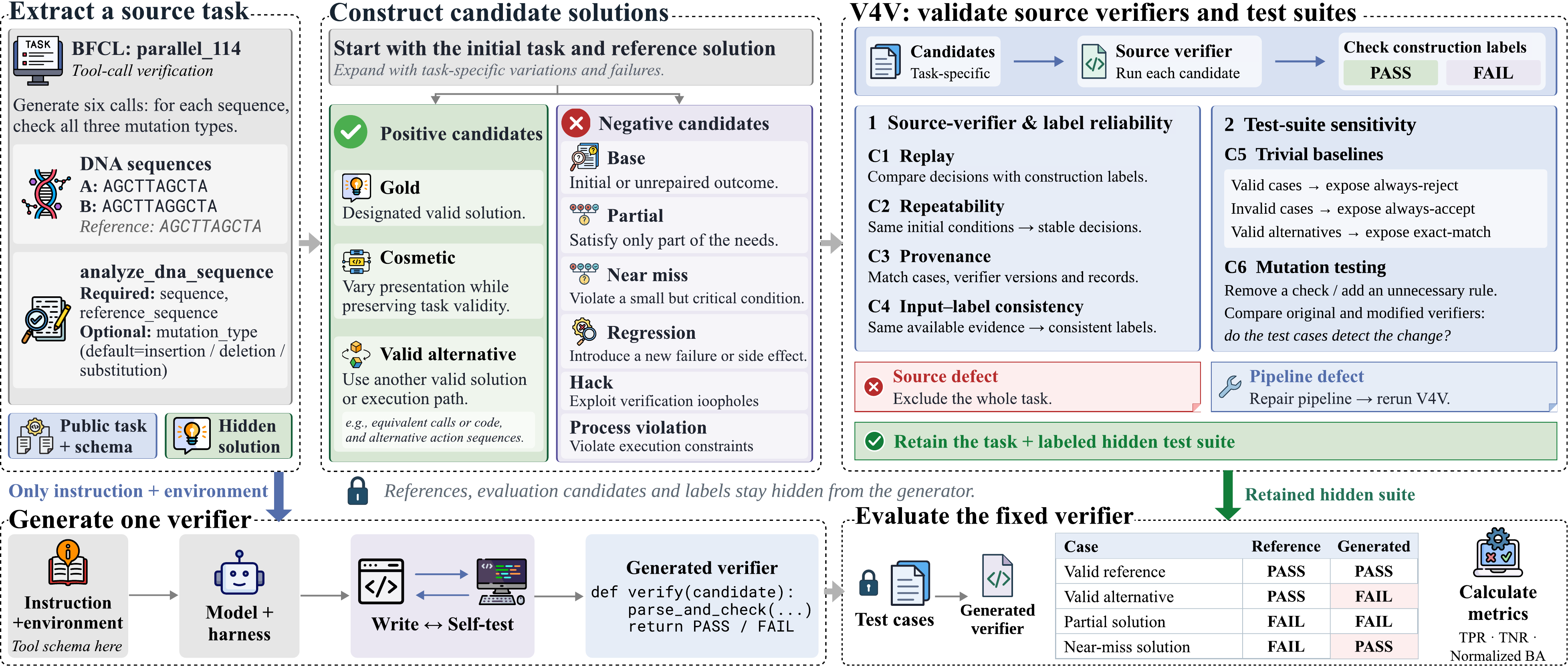}
\caption{\textbf{VerifierBench overview.} Source tasks and source verifiers are combined with constructed test suites; V4V screens both, and the retained source verifiers become reference verifiers. The agent receives the instruction and environment used for ordinary task solving and generates one verifier, which is fixed before it is scored on multiple positive and negative test cases.}
\label{fig:vb-overview}
\end{figure*}

\begin{figure*}[t]
\centering
\begingroup
\fontsize{9}{11}\selectfont
\setlength{\tabcolsep}{8pt}
\begin{tabularx}{\textwidth}{@{}>{\raggedright\arraybackslash}X>{\raggedright\arraybackslash}X@{}}
\toprule
\multicolumn{2}{@{}p{\dimexpr\textwidth-2\tabcolsep\relax}@{}}{\textbf{Task: credit-card retention.} Submit the requested card application and record the closure reason. Permitted tools include the read-only query \texttt{get\_user\_dispute\_history\_7291}.} \\
\multicolumn{2}{@{}p{\dimexpr\textwidth-2\tabcolsep\relax}@{}}{\textbf{Generated check:} The agent creates a tool allowlist that omits this query and rejects unlisted operations.} \\
\midrule
\textbf{Self-tests during generation} & \textbf{Hidden positive test cases} \\
The intended-positive candidate omits the dispute-history query. & Valid candidates include the permitted dispute-history query. \\
The positive self-test returns \texttt{FAIL}: replay is unavailable after budget exhaustion. & The verifier rejects the query as ``Unrelated or unsupported mutation.'' \\
The agent acknowledges that positive validation is incomplete. & All 9 positive cases receive \texttt{FAIL} without environment execution. \\
\midrule
\multicolumn{2}{@{}p{\textwidth}@{}}{An incomplete allowlist excludes a legal read; self-testing never confirms acceptance of a complete valid solution.} \\
\bottomrule
\end{tabularx}
\endgroup
\caption{\textbf{Overlooked task details lead to incorrect verification.} The agent's tool allowlist omits a permitted read-only query, causing valid solutions to be rejected. Its self-tests omit the query and never confirm acceptance. GPT-6-astra with Claude Code on $\tau^2$-bench banking task 044 rejects all 9 positive cases.}
\label{vb:rule-case}
\vspace{-6pt}
\end{figure*}

\section{VerifierBench}
\label{sec:benchmark}
\label{sec:task}

\subsection{Task Formulation}
\label{sec:generation}

For task $t$, an agent receives the instruction and environment used for ordinary task solving, together with the verifier's input and output conventions summarized in Appendix~\ref{app:contract}.
A \emph{harness}, the agent framework that manages model calls, tool use, and the generation loop, runs the agent until it delivers an executable verifier $\widehat V_t$; the reference verifier and hidden test cases are withheld.
During generation, the agent may run \emph{self-tests}: it writes candidate solutions of its own and checks that its draft verifier returns \texttt{PASS} for intended positives and \texttt{FAIL} for intended negatives.
Because the hidden test cases are withheld, self-tests are the agent's only evidence about how its verifier behaves.

After generation, we fix $\widehat V_t$ and run it on every test case in the task's suite $\mathcal C_t$.
Each test case holds one candidate solution and carries a reference label defined in Section~\ref{sec:validation}, \texttt{PASS} for a \emph{positive} case and \texttt{FAIL} for a \emph{negative} one, which the verifier never sees.

A tool-call verifier receives raw call text together with parsed function names and arguments; a state verifier receives the candidate's interaction record or, for AppWorld, its solution code; and a code verifier receives the candidate patch and a repository checkout.
For state and code tasks, both the generating agent and the delivered verifier can execute code in the task environment through a public interface, \texttt{run\_public}, for example to replay a candidate's actions or run a repository's tests.
Executions during generation are budgeted separately from those during evaluation, as documented in Appendix~\ref{app:setup}.

\subsection{Task Sources and Test Cases}
\label{sec:construction}
\label{sec:selection}
\label{sec:composition}
\label{sec:candidates}

We draw tasks from open-source agent benchmarks with executable evaluation procedures, each supplying an instruction, an environment, and a source verifier.
As summarized in the top of Table~\ref{vb:composition}, tool-call tasks check called functions and arguments, state tasks check interactions or the resulting environment state, and code tasks check proposed software changes.
Each test case receives a \emph{construction label}, \texttt{PASS} or \texttt{FAIL}, assigned from the task requirements and the change made to the solution before any verifier is run.
As listed in the bottom of Table~\ref{vb:composition}, test cases form four groups: the designated \emph{reference solution} (\texttt{gold}); \emph{valid variants}, which change presentation, implementation, or execution path while still satisfying every requirement; \emph{ordinary errors}, namely initial outcomes, partial solutions, near misses, and regressions; and \emph{attacks}, which exploit checking omissions, fabricate success signals, or interfere with verification.
Valid variants test whether a verifier accepts more than one expression of a correct solution.
Appendix~\ref{app:candidates} details the nine construction categories behind these groups.

\subsection{Screening with Verifier4Verifier}
\label{sec:validation}
\label{sec:information}

These test cases are useful only if their labels are reliable.
Construction labels and source verifiers can both be wrong: a constructed case may not change the solution as intended, and a source verifier may contain bugs.
V4V applies six checks to the constructed test cases and source verifiers:
\begin{enumerate}
\setlength{\itemsep}{0pt}\setlength{\parskip}{0pt}
    \item \textbf{Replay.} Run the source verifier on every test case and compare its decision with the construction label.
    \item \textbf{Repeatability.} Rerun it on the same case from the same reset state; the decision must not change.
    \item \textbf{Provenance.} Link every recorded decision to the test-case contents and verifier version that produced it.
    \item \textbf{Input-label consistency.} Within a task, cases that give the verifier identical information must not carry opposite labels.
    \item \textbf{Trivial baselines.} Verifiers that always accept, always reject, or accept only the reference solution must be exposed by negative cases, positive cases, and valid variants, respectively.
    \item \textbf{Mutation testing.} Modify one check in the source verifier \citep{demillo1978hints}; some test case must then receive different valid decisions from the two versions.
\end{enumerate}
We resolve any disagreement between a construction label and a source-verifier decision against the task requirements; agreement alone does not show that either is semantically correct.

As detailed in Appendix~\ref{app:v4v-exclusions}, confirmed source defects exclude 45 of the 267 initial tasks.
All 37 Terminal-Bench tasks are removed, 35 of them because its test-output parser counts skipped tests as passes; the others are five BFCL tasks whose answer keys conflict with default-argument rules and three SWE-bench tasks with parsing and empty-report errors.
Defects in our own evaluation pipeline are corrected without excluding tasks.
All 222 retained tasks pass the six checks, as reported in Appendix~\ref{app:v4v-retained}, and their source verifiers become the \emph{reference verifiers} $V_t^\star$.
Because replay agrees on every test case, each construction label equals the reference verifier's decision; we call this value the \emph{reference label} and score generated verifiers against it.

\begin{table*}[t]
\centering
\fontsize{10}{12}\selectfont
\setlength{\tabcolsep}{5pt}
\begin{tabularx}{\linewidth}{@{}l>{\raggedright\arraybackslash}Xrrr@{}}
\toprule
Task type & Source benchmarks & Tasks & \texttt{PASS} & \texttt{FAIL} \\
\midrule
Tool-call & BFCL v4 \citep{patil2025berkeley}, ACEBench \citep{chen2025acebench} & 95 & 734 & 2214 \\
State & $\tau^2$-bench \citep{barres2025tau}, AppWorld \citep{trivedi2024appworld}, AgentDojo \citep{debenedetti2024agentdojo} & 80 & 642 & 769 \\
Code & SWE-bench Verified \citep{jimenez2024swe} & 47 & 190 & 444 \\
\midrule
\textbf{Total} & & \textbf{222} & \textbf{1566} & \textbf{3427} \\
\bottomrule
\end{tabularx}

\vspace{6pt}
\begin{tabularx}{\linewidth}{@{}l c r >{\raggedright\arraybackslash}X@{}}
\toprule
Test-case group & Label & Cases & Content \\
\midrule
Reference solution & \texttt{PASS} & 222 & The designated solution (\texttt{gold}) \\
Valid variants & \texttt{PASS} & 1344 & Cosmetic changes (840) or alternative implementations or paths (504) \\
Ordinary errors & \texttt{FAIL} & 2629 & Initial, partial, near-miss, or regressing solutions \\
Attacks & \texttt{FAIL} & 798 & Exploited checking omissions, fabricated success signals, or interference \\
\bottomrule
\end{tabularx}
\caption{\textbf{Benchmark composition after V4V screening.} Top: tasks and reference labels by task type. Bottom: the four test-case groups. Appendix Tables~\ref{vb:source-composition} and~\ref{vb:taxonomy} give counts by source and by construction category.}
\label{vb:composition}
\vspace{-6pt}
\end{table*}

\subsection{Metrics}
\label{sec:metrics}

A verifier invocation yields a valid \texttt{PASS}, a valid \texttt{FAIL}, or no valid decision (a missing verifier, timeout, crash, or malformed output).
A verifier that catches its own internal error and returns \texttt{FAIL} still yields a valid decision, although it may not have completed its intended check.
For each task, we compute the fraction of positive cases receiving \texttt{PASS} and the fraction of negative cases receiving \texttt{FAIL}; the true positive rate (TPR) and true negative rate (TNR) average these fractions across tasks, in percent.
Our primary metric is normalized task-averaged balanced accuracy \citep{brodersen2010balanced}, which rescales $\mathrm{BA}=(\mathrm{TPR}+\mathrm{TNR})/2$ as
\begin{equation}
    \mathrm{BA}_{\rm norm}=2\,\mathrm{BA}-100=\mathrm{TPR}+\mathrm{TNR}-100.
    \label{eq:ba-norm}
\end{equation}
It weights tasks equally and the two classes equally within each task.
Constant valid decisions score 0 and perfect agreement scores 100; missing or invalid decisions stay in the denominators without credit.
We also report the attack false-positive rate FPR$_h$, the task-averaged fraction of attack cases that receive \texttt{PASS}. More results are in Appendices~\ref{app:metrics} and~\ref{app:extended}).

\section{Experiments}
\label{sec:experiments}

\subsection{Experimental Setup}
\label{sec:setup}

\paragraph{Models and harnesses.}
Our main comparison runs all 19 models in one common harness, Pi \citep{zechner_pi}.
To measure how much the harness matters, we add 17 alternative configurations.
They pair models with each family's \emph{native harness}, the vendor's own agent tool (Claude Code \citep{anthropic_claude_code}, Codex \citep{openai_codex}, or Gemini CLI \citep{google_gemini_cli}), with the third-party harnesses OpenClaw \citep{openclaw} and mini-swe \citep{yang2024sweagent}, and across families, as with GPT-6-astra in Claude Code.
Because harnesses differ in prompts, tools, and reasoning settings, cross-harness comparisons describe configurations rather than model capability alone.
Generation is limited to 5,400 seconds per task and 32,768 output tokens per reply, and every configuration is scored on all 222 tasks, as documented in Appendix~\ref{app:setup}.

\paragraph{Evidence sources.}
Our results draw on three sets of records.
The \emph{main release} holds the final judgments of all 36 configurations, and every reported score comes from it.
The \emph{archived records} are an earlier export with per-task judgments for nine Pi-harness models, whose aggregate counts match the main release, plus earlier verifier versions used in some appendix cases; the concentration analysis in Section~\ref{sec:errors} uses them.
The \emph{Astra trace batch} is a separate generation of GPT-6-astra verifiers under four conditions, Pi or Claude Code at LOW or HIGH reasoning effort, that retains generation traces.
It supports the effort comparison in Section~\ref{sec:harness-effort} and the trace analysis in Section~\ref{sec:analysis}, and its scores never replace those of the main release.

\subsection{Main Results}
\label{sec:overall}

Table~\ref{vb:main-results} ranks the 19 models.
BA$_{\rm norm}$ ranges from 10.6 to 62.1: GPT-6-astra ranks first, followed by Claude Opus 4.8 (54.0) and Kimi K3 (53.4), and no model approaches perfect agreement.

\begin{table}[t]
\centering
\fontsize{9}{11}\selectfont
\setlength{\tabcolsep}{3pt}
\renewcommand{\arraystretch}{1.02}
\begin{tabular*}{\columnwidth}{@{\extracolsep{\fill}}lrrrr@{}}
\toprule
Model & BA$_{\rm norm}$ & TPR & TNR & FPR$_h$ \\
\midrule
GPT-6-astra & \textbf{62.1} & 67.4 & 94.6 & 7.4 \\
Claude Opus 4.8 & 54.0 & 67.7 & 86.3 & 13.5 \\
Kimi K3 & 53.4 & \textbf{70.2} & 83.2 & 13.4 \\
GPT-5.6-sol & 52.9 & 56.8 & 96.0 & \textbf{0.8} \\
GPT-5.6-terra & 52.9 & 56.5 & \textbf{96.3} & 3.6 \\
GLM-5.2 & 48.8 & 68.1 & 80.8 & 12.1 \\
Doubao Seed 2.1 Pro & 46.8 & 57.0 & 89.9 & 11.4 \\
GPT-5.5 & 46.6 & 61.6 & 85.1 & 9.9 \\
DeepSeek V4 Pro & 44.8 & 65.5 & 79.2 & 10.1 \\
DeepSeek V4 Flash & 40.5 & 60.7 & 79.8 & 8.0 \\
Qwen3.7 Plus & 40.2 & 57.6 & 82.6 & 13.6 \\
Gemini 3.5 Flash & 35.7 & 52.4 & 83.3 & 13.2 \\
Kimi K2.6 & 35.2 & 52.5 & 82.7 & 4.6 \\
Doubao Seed 2.0 Code & 29.7 & 45.0 & 84.8 & 12.3 \\
Doubao Seed 2.0 Lite & 26.8 & 41.1 & 85.7 & 10.5 \\
Doubao Seed 2.0 Pro & 22.8 & 27.4 & 95.4 & 6.8 \\
GPT-5.1 & 22.2 & 35.2 & 87.1 & 9.7 \\
Gemini 2.5 Pro & 14.2 & 31.7 & 82.5 & 4.0 \\
Doubao Seed 2.0 Mini & 10.6 & 21.1 & 89.5 & 3.7 \\
\midrule
Constant / random & 0.0 & --- & --- & --- \\
\bottomrule
\end{tabular*}
\caption{\textbf{Main results with the common Pi harness}, on the same 222 tasks and ordered by BA$_{\rm norm}=\mathrm{TPR}+\mathrm{TNR}-100$. Rates are percentages, rounded independently; bold marks the best value per column (lowest for FPR$_h$). Constant baselines return valid decisions throughout, and random scores are expectations. Table~\ref{vb:native-harness} and Appendix Table~\ref{vb:alt-results} cover the 17 alternative configurations.}
\label{vb:main-results}
\end{table}

\begin{figure*}[!t]
\centering
\includegraphics[width=0.71\textwidth]{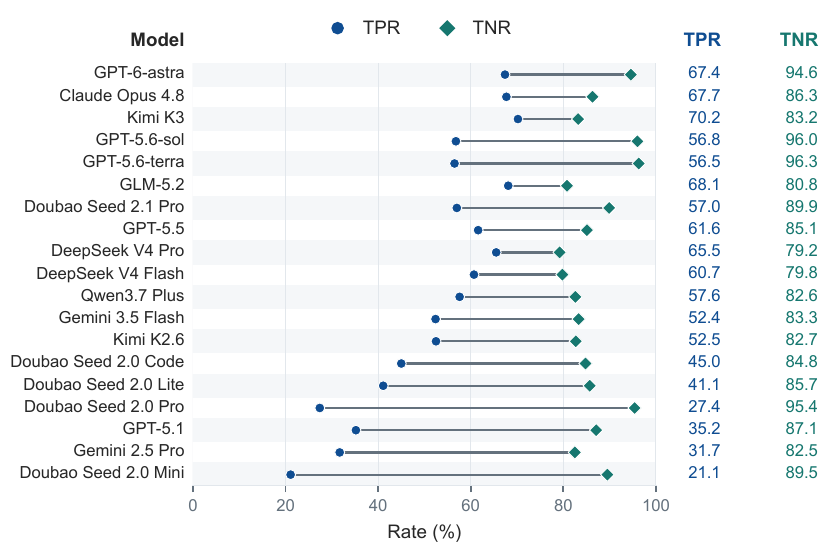}
\caption{\textbf{Acceptance versus rejection.} TPR and TNR for the 19 models with the Pi harness, ordered by descending BA$_{\rm norm}$. Every model accepts positive cases (TPR) less reliably than it rejects negative ones (TNR).}
\label{vb:results}
\label{vb:tpr-tnr}
\end{figure*}

Accepting valid solutions is the shared weakness, as shown in Figure~\ref{vb:results}.
Every model has lower TPR than TNR: TPR ranges from 21.1\% to 70.2\%, whereas TNR ranges from 79.2\% to 96.3\%.
As Appendix Table~\ref{vb:positive-diagnostics} reports, explicit \texttt{FAIL}, rather than a missing decision, accounts for 57.1\%--100\% of each model's unaccepted positive cases, so the gap mainly reflects active rejection of valid solutions.
As detailed in Appendix~\ref{app:task-types}, the gap is widest on state and code tasks, where TPR averaged over models is 27.7\% and 40.2\%, compared with 79.2\% on tool-call tasks.

Low attack acceptance does not by itself indicate a reliable verifier.
As Appendix~\ref{app:task-types} further shows, on state tasks GPT-5.6-sol has an FPR$_h$ of 0.0\% but accepts only 23.0\% of positive cases, and on code tasks 17 models have an FPR$_h$ of 0.0\% at the reported precision while their BA$_{\rm norm}$ ranges from $-25.9$ to 76.9.

\subsection{Effects of Harness and Reasoning Effort}
\label{sec:harness-effort}

\paragraph{Native harnesses.}
As Table~\ref{vb:native-harness} shows, native harnesses change scores in both directions.
Claude Code raises Claude Opus 4.8 by 12.2 points to 66.2, the best of all 36 configurations, and Gemini CLI raises Gemini 3.5 Flash by 15.0 points, both through higher TPR and higher TNR.
In contrast, Gemini CLI lowers Gemini 2.5 Pro by 9.2 points, and Codex lowers all four GPT models by 4.1--26.6 points.
The harnesses also run with different reasoning settings, as noted in Appendix~\ref{app:setup}: retained Opus/Claude Code traces record high effort with thinking enabled, whereas historical Pi traces that expose a thinking setting record it as off.

Paired code and trace inspection reveals concrete failures behind these differences.
Some GPT-5.5/Codex verifiers require particular patch contents or repair locations and reject valid fixes before any functional check.
On a booking-cancellation task, the GPT-5.6-sol/Codex verifier additionally requires the raw call text to parse as JSON, and its self-test supplies only the parsed calls, so it never exercises this check.
As Appendix Figure~\ref{fig:demo-sol} illustrates, the verifier accepts none of the 7 positive cases, whereas the Pi verifier accepts all 7; both reject all 20 negative cases.
Conversely, on retail/49 the Claude Opus 4.8/Claude Code agent executes the target exchange to obtain the expected environment state, and its verifier accepts all 9 positive cases.
As Appendix Figure~\ref{fig:demo-opus} shows, the Pi verifier instead relies on replaying an expected exchange that was never successfully executed during generation and accepts none; both reject all 8 negative cases.
Selected Gemini 3.5 Flash/Gemini CLI verifiers likewise avoid implementation errors that prevent valid decisions.

\paragraph{Other alternative harnesses.}
As Appendix Tables~\ref{vb:harness} and~\ref{vb:alt-results} show, no harness improves every model it runs.
Codex lowers all five models it runs, including Qwen3.7 Plus ($-14.1$ points).
OpenClaw and mini-swe lower GPT-6-astra and GPT-5.6-sol by 7.2--17.8 points but change Qwen3.7 Plus by at most 0.3 points, and Claude Code raises Qwen3.7 Plus by 8.7 points but lowers GPT-5.6-sol by 6.5.
The changes also arise in different ways.
For Qwen3.7 Plus, Codex raises TNR from 82.6\% to 92.1\% but lowers TPR from 57.6\% to 34.0\%, whereas Claude Code raises both rates for Claude Opus 4.8, so harness effects cannot be reduced to a fixed trade-off between acceptance and rejection.

\begin{table}[t]
\centering
\fontsize{9}{11}\selectfont
\setlength{\tabcolsep}{2.5pt}
\renewcommand{\arraystretch}{1.05}
\begin{tabular*}{\columnwidth}{@{\extracolsep{\fill}}llrrr@{}}
\toprule
Model & Native & Pi & Native & $\Delta$ \\
\midrule
Claude Opus 4.8 & Claude Code & 54.0 & \textbf{66.2} & $+12.2$ \\
GPT-5.5 & Codex & \textbf{46.6} & 30.3 & $-16.4$ \\
GPT-5.6-sol & Codex & \textbf{52.9} & 26.2 & $-26.6$ \\
GPT-5.6-terra & Codex & \textbf{52.9} & 31.0 & $-21.8$ \\
GPT-6-astra & Codex & \textbf{62.1} & 58.0 & $-4.1$ \\
Gemini 3.5 Flash & Gemini CLI & 35.7 & \textbf{50.7} & $+15.0$ \\
Gemini 2.5 Pro & Gemini CLI & \textbf{14.2} & 5.0 & $-9.2$ \\
\bottomrule
\end{tabular*}
\caption{\textbf{Native harness comparison.} BA$_{\rm norm}$ on the same 222 tasks with Pi and with each family's native harness. Bold marks the higher score in each pair; $\Delta$ is native minus Pi, rounded independently.}
\label{vb:native-harness}
\end{table}

\paragraph{Reasoning effort.}
As Appendix Table~\ref{vb:reasoning} reports, reasoning effort, by contrast, barely changes aggregate scores in the Astra trace batch.
With Pi, HIGH instead of LOW effort raises TPR by 2.8 points but lowers TNR by 2.3 and raises FPR$_h$ by 2.2, for a net BA$_{\rm norm}$ change of $+0.5$; with Claude Code, all three move by less than one point, and BA$_{\rm norm}$ changes by $-1.3$.
These small totals hide opposite changes by task type, broken down in Appendix Table~\ref{vb:effort-types}: HIGH effort improves code tasks with both harnesses ($+3.8$ and $+1.4$ points) but worsens state tasks ($-5.4$ and $-3.6$), while tool-call tasks change by $+3.9$ with Pi and $-0.6$ with Claude Code.
Higher acceptance does not imply better discrimination: on Pi tool-call tasks, HIGH effort accepts 63 more positive cases without newly rejecting any, but it also introduces 128 false acceptances while correcting three.

\paragraph{Execution budget.}
The clearest effect of effort is on the generation budget of 20 \texttt{run\_public} executions per task.
As Appendix Table~\ref{vb:effort-budget} reports, among the 127 state and code tasks, runs that exhaust this budget rise from 4 to 77 with Pi and from 21 to 98 with Claude Code.
On the 32 $\tau^2$-bench tasks, runs that exhaust it before first writing the verifier rise from 0 to 30 with Pi and from 7 to 20 with Claude Code.

\subsection{Where False Rejections Concentrate}
\label{sec:errors}

These rates do not show whether false rejections are spread across tasks or repeated within the same task.
We call a verifier \emph{all-reject} if it returns a valid \texttt{FAIL} on every test case of its task, positive and negative alike.
On that task its TPR is 0 and its TNR is 100, so it contributes zero to BA$_{\rm norm}$, like a constant decision.

All-reject verifiers produce most false rejections.
As Appendix Figure~\ref{vb:concentration} shows, in the archived records they contribute 87.9--96.0\% of each model's positive \texttt{FAIL} decisions, whether test cases or tasks are weighted equally.
The Astra trace batch reproduces the pattern, as reported in Appendix Table~\ref{vb:final-concentration}: in each condition, 68--75 of the 222 verifiers are all-reject and contribute 93.3--95.5\% of false rejections.

The concentration of false rejections identifies where failures occur, but not why.
BA$_{\rm norm}$ penalizes these failures but does not identify the checks responsible for them.
We therefore inspect selected verifiers and their generation traces in Section~\ref{sec:analysis}.

\section{Why Generated Verifiers Reject Valid Solutions}
\label{sec:analysis}

Section~\ref{sec:errors} showed that false rejections concentrate in all-reject verifiers.
We now ask which check rejects the valid solutions, in Section~\ref{sec:rules}, and why the agent's self-tests did not expose it, in Sections~\ref{sec:selftests} and~\ref{sec:incomplete}.
We read generated code and generation traces from the Astra trace batch.
For each rejected positive case, we compare the rejected property of the solution with the public task requirements and with the generated check, and attribute the rejection to that check only when the requirements permit the property.
Because the verifiers are selected to explain observed failures, the analysis identifies mechanisms rather than their prevalence.

\subsection{Unjustified Checks}
\label{sec:rules}
\label{sec:cases}

Table~\ref{vb:rule-defects} lists 16 verifiers from 12 tasks whose rejections we attribute to an \emph{unjustified check}: a condition that the instruction, tool definitions, and environment rules do not require, and that rejects solutions they permit.
One such check can reject several positive cases of the same task.

\begin{table}[t]
\centering
\fontsize{10}{12}\selectfont
\setlength{\tabcolsep}{3pt}
\begin{tabularx}{\columnwidth}{@{}>{\raggedright\arraybackslash}Xrrr@{}}
\toprule
Unjustified check & Tasks & Verifiers & \shortstack{Positive\\\texttt{FAIL}} \\
\midrule
\multicolumn{4}{@{}l}{\emph{Self-tests used only allowed solutions}} \\
Overbroad source-file protection & 2 & 3 & 12 \\
Rejects permitted representations & 3 & 5 & 35 \\
\addlinespace[2pt]
\emph{Subtotal} & 5 & 8 & 47 \\
\midrule
\multicolumn{4}{@{}l}{\emph{No positive self-test passed}} \\
Rejects authorized user-tool calls & 3 & 3 & 27 \\
Omits permitted read-only tools & 2 & 2 & 18 \\
Incomplete category vocabulary & 2 & 3 & 27 \\
\addlinespace[2pt]
\emph{Subtotal} & 7 & 8 & 72 \\
\midrule
\textbf{Total} & \textbf{12} & \textbf{16} & \textbf{119} \\
\bottomrule
\end{tabularx}
\caption{\textbf{Unjustified checks in selected Astra verifiers.} Lower block: no positive self-test returned a complete \texttt{PASS} during generation. Upper block: the remaining verifiers, whose self-tests used only solutions the check allows. Each verifier is a task--condition pair; the 119 positive \texttt{FAIL} decisions cover 90 distinct test cases. The review is targeted and does not estimate prevalence.}
\label{vb:rule-defects}
\end{table}

\paragraph{Tool calls.}
In $\tau^2$-bench banking task 044, shown in Figure~\ref{vb:rule-case}, the Claude Code/LOW verifier uses a tool allowlist to exclude unrelated mutations but omits \nolinkurl{get_user_dispute_history_7291}, a permitted read-only query included in the public tool list.
It treats this legal query as an unsupported mutation and rejects all nine positive cases before environment execution.
The Pi/LOW verifier for task 045 omits the same tool and also rejects all nine positives.
Three other Pi/LOW verifiers (tasks 037, 040, and 041) reject authorized user-side calls or lack a branch for them, rejecting another 27 positives without environment execution.

\paragraph{Source files.}
Both Claude Code verifiers for Matplotlib-14623 exclude legal changes under \nolinkurl{lib/mpl_toolkits/}, including the 3D axes implementation, and each rejects all four positive cases.

\paragraph{Category vocabulary.}
In retail/34, the Claude Code/HIGH verifier identifies the task's three office items with a keyword list that omits \texttt{Headphones}, counts only two office items, and rejects all nine positive cases, as Appendix Figure~\ref{fig:demo-retail} illustrates; two other HIGH-effort verifiers, for retail/33 and retail/34, share the omission.

\paragraph{Representations.}
As detailed in Appendix~\ref{app:rule-cases}, the remaining verifiers accept only some permitted representations of a value: an ISO-only currency format, numeric-only device identifiers, or incomplete decoding of TeX spacing.

\subsection{Self-Tests That Confirm Without Challenging}
\label{sec:selftests}

If these checks are visible in the generated code, why did self-testing not catch them?
Table~\ref{vb:rule-defects} splits the 16 verifiers by self-test outcome.
For the eight in the upper block, the self-tests used only solutions that the unjustified check already allows.
In Django-11734, the positive self-test patches \texttt{query.py}; the agent repairs subprocess compatibility, patch formatting, and environment resets until its positive and negative self-tests return \texttt{PASS} and \texttt{FAIL} as intended, yet no self-test edits an \texttt{\_\_init\_\_.py} file.
The passing Matplotlib self-tests modify other locations under \nolinkurl{lib/matplotlib/}, such as \nolinkurl{axes/_base.py} and \texttt{ticker.py}, and the representation self-tests use only representations the checks accept.

We call this \emph{confirmatory self-testing}: the tests confirm that the verifier accepts a chosen solution but never ask whether its other conditions follow from the task.
More tests at the same permitted locations would not expose the restriction, and neither would negative self-tests: an invalid patch may violate a genuine requirement while also tripping the unjustified check, so rejecting it correctly does not justify the check.
As Appendix Figure~\ref{fig:demo-dna} shows, a Qwen3.7 Plus verifier exhibits the same pattern outside the Astra batch: it treats an omitted optional argument as empty instead of applying its documented default, and every positive self-test supplies that argument explicitly.

\subsection{Self-Tests That Never Confirm Acceptance}
\label{sec:incomplete}
\label{sec:reasoning}
\label{sec:exploration}

The eight verifiers in the lower block of Table~\ref{vb:rule-defects} are the banking and retail verifiers of Section~\ref{sec:rules}.
Their agents ran intended-positive self-tests but never obtained a complete positive \texttt{PASS}, as each agent's final response acknowledges, so these runs do not establish acceptance even of the agent's own intended solution.
In the running example of Figure~\ref{vb:rule-case}, the intended-positive self-test omits the permitted dispute-history query and fails when replay is unavailable after budget exhaustion; the tool restriction remains untested.
At evaluation, the eight verifiers reject all 72 of their positive judgments without environment execution.

Generation-time testing is constrained more broadly.
Exploration and self-testing share one execution budget, which most HIGH-effort state and code runs exhaust as described in Section~\ref{sec:harness-effort}, and in two appendix traces shown in Appendix Figures~\ref{fig:demo-sol} and~\ref{fig:demo-opus} local execution fails because \texttt{unshare} is unavailable.
These limits do not directly cause hidden rejections, since evaluation has its own budget, but they can leave a verifier's checks unresolved when generation ends.

\section{Implications}
\label{sec:implications}

Self-testing a generated verifier serves two goals: confirming that it accepts a chosen solution, and checking that it does not reject other solutions the task allows.
The second requires tracing every acceptance condition to the instruction, tool definitions, or environment rules, and testing it with cases derived from those requirements.
One test varies a permitted property while keeping the task satisfied, for example editing \texttt{fields/\_\_init\_\_.py} instead of \texttt{query.py}, omitting a defaulted argument, or using an authorized user-side call; another violates one required condition.
Examples drawn from solutions a check already allows cannot challenge that check, and the Django configurations show that permitting legal source changes need not weaken rejection of tested violations.

Testing must also be completed.
Generation should reserve execution budget for testing the final verifier and should distinguish an execution failure from a failed requirement: a blocked test leaves its check unresolved, and repairing the execution does not validate checks the test never exercises.
We do not evaluate the effectiveness of these directions.

\section{Related Work}
\label{sec:related}

\textbf{Executable verifier generation.} nl2postcond \citep{endres2024can} and CodeSpecBench \citep{chen2026codespecbench} generate executable specifications and evaluate them on valid and faulty behavior; CheckerBench \citep{he2026checkerbench} studies agent-generated static-analysis checkers. We evaluate task-completion verifiers for tool-call, state, and code tasks, generated from the instruction and environment available for task solving.
\textbf{Test-oracle and test generation.} TOGA \citep{dinella2022toga} generates test assertions, TestGenEval \citep{jain2025testgeneval} evaluates unit-test generation, and SWT-Bench \citep{mundler2024swt} generates issue-reproducing tests that also assess candidate patches. Our suites contain multiple valid variants and violations per task, and our trace analysis examines whether self-tests challenge unjustified checks.
\textbf{Automated training-data construction.} Absolute Zero \citep{zhao2026absolute} learns from self-proposed tasks with execution-based verification. Self-Challenging Language Model Agents \citep{zhou2026self} jointly generates tasks, verification functions, and test examples for reinforcement learning. We hold tasks fixed to evaluate verifier generation in isolation, and identify errors that could affect which solutions enter training data.

\section{Conclusion}
\label{sec:conclusion}

VerifierBench evaluates whether agents can write verifiers from the information available for solving a task.
Across all 36 configurations, generated verifiers accept valid solutions less reliably than they reject invalid ones, and in the analyzed records most false rejections come from verifiers that reject every case of their task.
Code and trace inspection links these verifiers to unjustified checks and to self-tests that either confirm only solutions those checks allow or never confirm acceptance at all.
Verifier generation therefore needs tests that challenge each acceptance condition against the task requirements; otherwise, where generated verifiers select training data, as in data-oriented RSI, such checks could repeatedly exclude legitimate solution strategies.

\clearpage
\section*{Limitations}

Scores measure agreement with retained reference verifiers on finite, constructed test suites.
As detailed in Appendix~\ref{app:qualification}, V4V excludes 45 tasks with confirmed source defects, but agreement between construction labels and reference decisions does not certify that every label is semantically correct.
The 36 configurations differ in harness prompts, tools, and reasoning settings, so cross-harness differences describe configurations and do not isolate any harness component; rankings are point estimates without repeated generations or significance tests.
The concentration analysis uses archived per-task records for nine models and the Astra trace batch for one model, rather than final per-task records for all 19 models.
The code and trace analysis covers verifiers selected to explain observed failures, mostly from GPT-6-astra.
It attributes rejections by inspecting source code rather than by intervention experiments, and it reads self-test outcomes from traces without reconstructing the verifier version at each self-test; it therefore identifies mechanisms but not their prevalence.

\section*{AI Assistance Disclosure}
We used AI assistants for manuscript refinement, code structuring, and debugging.

\bibliography{custom}

@article{endres2024can,
  title={Can large language models transform natural language intent into formal method postconditions?},
  author={Endres, Madeline and Fakhoury, Sarah and Chakraborty, Saikat and Lahiri, Shuvendu K},
  journal={Proceedings of the ACM on Software Engineering},
  volume={1},
  number={FSE},
  pages={1889--1912},
  year={2024},
  publisher={ACM New York, NY, USA}
}

@misc{zechner_pi,
  author       = {Zechner, Mario},
  title        = {{pi}: Coding Agent},
  year         = {2025},
  howpublished = {Software repository},
  url          = {https://github.com/earendil-works/pi},
  note         = {Accessed: 2026-10-11}
}

@misc{openai_codex,
  author       = {{OpenAI}},
  title        = {{Codex CLI}},
  year         = {2025},
  howpublished = {Software repository},
  url          = {https://github.com/openai/codex},
  note         = {Accessed: 2026-10-11}
}

@misc{anthropic_claude_code,
  author       = {{Anthropic}},
  title        = {{Claude Code}},
  year         = {2025},
  howpublished = {Software repository},
  url          = {https://github.com/anthropics/claude-code},
  note         = {Accessed: 2026-10-11}
}

@misc{google_gemini_cli,
  author       = {{Google}},
  title        = {{Gemini CLI}},
  year         = {2025},
  howpublished = {Software repository},
  url          = {https://github.com/google-gemini/gemini-cli},
  note         = {Accessed: 2026-10-11}
}

@misc{openclaw,
  author       = {{OpenClaw Foundation}},
  title        = {{OpenClaw}},
  year         = {2026},
  howpublished = {Software repository},
  url          = {https://github.com/openclaw/openclaw},
  note         = {Accessed: 2026-10-11}
}

@inproceedings{yang2024sweagent,
  author    = {John Yang and Carlos E Jimenez and Alexander Wettig
               and Kilian Lieret and Shunyu Yao
               and Karthik R Narasimhan and Ofir Press},
  title     = {{SWE}-agent: Agent-Computer Interfaces Enable
               Automated Software Engineering},
  booktitle = {The Thirty-eighth Annual Conference on
               Neural Information Processing Systems},
  year      = {2024},
  url       = {https://arxiv.org/abs/2405.15793}
}

@article{chen2026codespecbench,
  title={CodeSpecBench: Benchmarking LLMs for Executable Behavioral Specification Generation},
  author={Chen, Zaoyu and Dai, Jianbo and Zhu, Boyu and Wang, Jingdong and Wang, Huiming and Xu, Xin and Yuan, Haoyang and Guo, Zhijiang and Wu, Xiao-Ming},
  journal={arXiv preprint arXiv:2604.12268},
  year={2026}
}

@article{he2026checkerbench,
  title={CheckerBench: Can Long-Horizon Agents Synthesize Static-Analysis Checkers?},
  author={He, Hang and Wang, Li and Chen, Hao and Shao, Yuchen and Shi, Yuling and Wang, Lisheng and Liu, Peiyang and Lin, Goose and Wang, Zaiyuan and Sun, Haiying and others},
  journal={arXiv preprint arXiv:2610.07557},
  year={2026}
}

@inproceedings{jimenez2024swe,
  title={Swe-bench: Can language models resolve real-world github issues?},
  author={Jimenez, Carlos E and Yang, John and Wettig, Alexander and Yao, Shunyu and Pei, Kexin and Press, Ofir and Narasimhan, Karthik},
  booktitle={International Conference on Learning Representations},
  volume={2024},
  pages={54107--54157},
  year={2024}
}

@article{debenedetti2024agentdojo,
  title={Agentdojo: A dynamic environment to evaluate prompt injection attacks and defenses for llm agents},
  author={Debenedetti, Edoardo and Zhang, Jie and Balunovic, Mislav and Beurer-Kellner, Luca and Fischer, Marc and Tram{\`e}r, Florian},
  journal={Advances in neural information processing systems},
  volume={37},
  pages={82895--82920},
  year={2024}
}

@inproceedings{yin2025godel,
  title={G{\"o}del agent: A self-referential agent framework for recursively self-improvement},
  author={Yin, Xunjian and Wang, Xinyi and Pan, Liangming and Lin, Li and Wan, Xiaojun and Wang, William Yang},
  booktitle={Proceedings of the 63rd Annual Meeting of the Association for Computational Linguistics (Volume 1: Long Papers)},
  pages={27890--27913},
  year={2025}
}

@article{min2026tracedance,
  title={TraceDance: An Automated System for Building Agent Behavior Benchmarks from Real-World Agent Deployment Traces},
  author={Min, Dehai and Zhang, Daoan and Zeng, Yiming and Zhang, Huayi and Chen, Ziyi and Zhang, Yan and Bai, Qinbo and Chao, Mengyuan and Ning, Jing and Hua, Qiyue and others},
  journal={arXiv preprint arXiv:2609.33295},
  year={2026}
}

@inproceedings{zhang2025generative,
  title={Generative verifiers: Reward modeling as next-token prediction},
  author={Zhang, Lunjun and Hosseini, Arian and Bansal, Hritik and Kazemi, Seyed Mehran and Kumar, Aviral and Agarwal, Rishabh},
  booktitle={International Conference on Learning Representations},
  volume={2025},
  pages={12476--12505},
  year={2025}
}

@inproceedings{brodersen2010balanced,
  title={The balanced accuracy and its posterior distribution},
  author={Brodersen, Kay Henning and Ong, Cheng Soon and Stephan, Klaas Enno and Buhmann, Joachim M},
  booktitle={2010 20th international conference on pattern recognition},
  pages={3121--3124},
  year={2010},
  organization={IEEE}
}

@inproceedings{dinella2022toga,
  title={Toga: A neural method for test oracle generation},
  author={Dinella, Elizabeth and Ryan, Gabriel and Mytkowicz, Todd and Lahiri, Shuvendu K},
  booktitle={Proceedings of the 44th International Conference on Software Engineering},
  pages={2130--2141},
  year={2022}
}

@article{demillo1978hints,
  title={Hints on test data selection: Help for the practicing programmer},
  author={DeMillo, Richard A and Lipton, Richard J and Sayward, Frederick G},
  journal={Computer},
  volume={11},
  number={4},
  pages={34--41},
  year={1978},
  publisher={IEEE}
}

@article{zhou2026self,
  title={Self-challenging language model agents},
  author={Zhou, Yifei and Levine, Sergey and Weston, Jason and Li, Xian and Sukhbaatar, Sainbayar},
  journal={Advances in Neural Information Processing Systems},
  volume={38},
  pages={113959--113991},
  year={2026}
}

@article{zhao2026absolute,
  title={Absolute zero: Reinforced self-play reasoning with zero data},
  author={Zhao, Andrew and Wu, Yiran and Wu, Tong and Xu, Quentin and Yue, Yang and Lin, Matthieu and Wang, Shenzhi and Wu, Qingyun and Zheng, Zilong and Huang, Gao},
  journal={Advances in Neural Information Processing Systems},
  volume={38},
  pages={105816--105879},
  year={2026}
}

@article{mundler2024swt,
  title={Swt-bench: Testing and validating real-world bug-fixes with code agents},
  author={M{\"u}ndler, Niels and M{\"u}ller, Mark N and He, Jingxuan and Vechev, Martin},
  journal={Advances in Neural Information Processing Systems},
  volume={37},
  pages={81857--81887},
  year={2024}
}

@inproceedings{jain2025testgeneval,
  title={Testgeneval: A real world unit test generation and test completion benchmark},
  author={Jain, Kush and Synnaeve, Gabriel and Rozi{\`e}re, Baptiste},
  booktitle={International Conference on Learning Representations},
  volume={2025},
  pages={14947--14999},
  year={2025}
}

@inproceedings{patil2025berkeley,
  title={The berkeley function calling leaderboard (bfcl): From tool use to agentic evaluation of large language models},
  author={Patil, Shishir G and Mao, Huanzhi and Yan, Fanjia and Ji, Charlie Cheng-Jie and Suresh, Vishnu and Stoica, Ion and Gonzalez, Joseph E},
  booktitle={Forty-second International Conference on Machine Learning},
  year={2025}
}

@inproceedings{chen2025acebench,
  title={ACEBench: A Comprehensive Evaluation of LLM Tool Usage.},
  author={Chen, Chen and Hao, Xinlong and Liu, Weiwen and Huang, Xu and Zeng, Xingshan and Yu, Shuai and Li, Dexun and Huang, Yuefeng and Liu, Xiangcheng and Wang, Xinzhi and others},
  booktitle={EMNLP (Findings)},
  pages={12970--12998},
  year={2025}
}

@article{barres2025tau,
  title={{$\tau^2$}-bench: Evaluating Conversational Agents in a Dual-Control Environment},
  author={Barres, Victor and Dong, Honghua and Ray, Soham and Si, Xujie and Narasimhan, Karthik},
  journal={arXiv preprint arXiv:2506.07982},
  year={2025}
}

@inproceedings{trivedi2024appworld,
  title={Appworld: A controllable world of apps and people for benchmarking interactive coding agents},
  author={Trivedi, Harsh and Khot, Tushar and Hartmann, Mareike and Manku, Ruskin and Dong, Vinty and Li, Edward and Gupta, Shashank and Sabharwal, Ashish and Balasubramanian, Niranjan},
  booktitle={Proceedings of the 62nd Annual Meeting of the Association for Computational Linguistics (Volume 1: Long Papers)},
  pages={16022--16076},
  year={2024}
}

\clearpage
\appendix
\raggedbottom
\setcounter{topnumber}{3}
\setcounter{bottomnumber}{2}
\setcounter{totalnumber}{5}
\setcounter{dbltopnumber}{3}
\renewcommand{\topfraction}{0.90}
\renewcommand{\bottomfraction}{0.85}
\renewcommand{\textfraction}{0.08}
\renewcommand{\floatpagefraction}{0.75}
\renewcommand{\dbltopfraction}{0.90}
\renewcommand{\dblfloatpagefraction}{0.75}
\setlength{\floatsep}{14pt plus 2pt minus 2pt}
\setlength{\textfloatsep}{16pt plus 2pt minus 2pt}
\setlength{\intextsep}{12pt plus 2pt minus 2pt}
\setlength{\dblfloatsep}{16pt plus 2pt minus 2pt}
\setlength{\dbltextfloatsep}{16pt plus 2pt minus 2pt}
\makeatletter
\setlength{\@fptop}{0pt}
\setlength{\@fpsep}{18pt}
\setlength{\@fpbot}{0pt plus 1fil}
\setlength{\@dblfptop}{0pt}
\setlength{\@dblfpsep}{18pt}
\setlength{\@dblfpbot}{0pt plus 1fil}
\makeatother

\section{Task Contract}
\label{app:contract}

Table~\ref{vb:contract} summarizes, for a single task, what the model receives, what it must produce, and how the resulting verifier is scored.

\par\medskip
\noindent\begin{minipage}{\columnwidth}
\makeatletter\def\@captype{table}\makeatother
\centering
\fontsize{10}{12}\selectfont
\setlength{\tabcolsep}{4pt}
\renewcommand{\arraystretch}{1.15}
\begin{tabularx}{\linewidth}{@{}l>{\raggedright\arraybackslash}X@{}}
\toprule
Element & Content \\
\midrule
Model input & Task instruction and environment, identical to the solving setting; the reference implementation and hidden test cases are withheld. \\
Model output & One executable verifier $\widehat{V}_t$ for task $t$. \\
Test suite & Positive cases (the reference solution and valid variants) and negative cases (ordinary errors and attacks), each carrying a reference label. \\
Reference & A fixed reference verifier $V_t^{\star}$ retained after V4V screening; its decision on every test case equals the reference label. \\
Scoring & For each hidden case, compare $\widehat{V}_t$'s \texttt{PASS}/\texttt{FAIL} decision against the reference label. \\
Metrics & Task-weighted TPR and TNR, summarized by BA$_{\rm norm}$. \\
\bottomrule
\end{tabularx}
\caption{Task contract for verifier generation and evaluation.}
\label{vb:contract}
\end{minipage}
\par\medskip

\section{Task Sources and Reference Validation}
\label{app:qualification}

Table~\ref{vb:source-composition} lists the 222 tasks retained from the initial collection of 267.
The initial collection also includes Terminal-Bench, but every Terminal-Bench task is removed during screening, so no Terminal-Bench task reaches the frozen benchmark; the retained tasks therefore form seven source subsets drawn from six benchmarks, with AgentDojo contributing separate user-task and injection subsets.
All main-table configurations use this complete task set.
The 45 exclusions concern reported defects in the source materials, whereas defects introduced by the evaluation pipeline are corrected without removing the task.
\begin{table*}[!tp]
\centering
\fontsize{10}{12}\selectfont
\setlength{\tabcolsep}{3.5pt}
\renewcommand{\arraystretch}{1.10}

\begin{tabularx}{\textwidth}{@{}>{\raggedright\arraybackslash}X l r r r r r@{}}
\toprule
Source & Type & Tasks & \texttt{PASS} & \texttt{FAIL} & Variants & Attack \\
\midrule
BFCL v4 & Tool-call & 45 & 390 & 1247 & 345 & 270 \\
ACEBench & Tool-call & 50 & 344 & 967 & 294 & 200 \\
\addlinespace[2pt]
\multicolumn{2}{@{}l}{\emph{Tool-call subtotal}} & 95 & 734 & 2214 & 639 & 470 \\
\midrule
$\tau^2$-bench & State & 32 & 287 & 303 & 255 & 96 \\
AppWorld & State & 28 & 216 & 278 & 188 & 56 \\
AgentDojo injection & State & 10 & 60 & 86 & 50 & 42 \\
AgentDojo user & State & 10 & 79 & 102 & 69 & 40 \\
\addlinespace[2pt]
\multicolumn{2}{@{}l}{\emph{State subtotal}} & 80 & 642 & 769 & 562 & 234 \\
\midrule
SWE-bench Verified & Code & 47 & 190 & 444 & 143 & 94 \\
\midrule
\multicolumn{2}{@{}l}{\textbf{Total}} & \textbf{222} & \textbf{1566} & \textbf{3427} & \textbf{1344} & \textbf{798} \\
\bottomrule
\end{tabularx}
\par\smallskip
\begin{minipage}{\linewidth}\fontsize{10}{12}\selectfont
All configurations are scored on the full task inventory. Missing submissions and invalid decisions remain in the scoring denominator.
\end{minipage}
\caption{\textbf{Task sources.} Inventory of the 222 tasks. \texttt{PASS} and \texttt{FAIL} are reference labels. Variants counts valid variants (cosmetic changes and alternative solutions); Attack combines \texttt{hack} and \texttt{process\_violation}. These are subsets of \texttt{PASS} and \texttt{FAIL}, respectively.}
\label{vb:source-composition}
\end{table*}

The following results describe the frozen task set used in the main comparison.\footnote{\raggedright Registry: \texttt{full-upstream-}\allowbreak\texttt{exclusion-20260919-v3}. Data revision: \texttt{swe-goldholes-native-}\allowbreak\texttt{20260919}.}
Test cases and their construction labels are created before screening.
Replay compares the source verifier's output with those labels; repeatability compares repeated executions with one another.

\paragraph{Licensing and redistribution.}
Original VerifierBench scripts and annotations use Apache 2.0; upstream-derived content retains its source terms.
ACEBench, $\tau^2$-bench, AgentDojo, and the SWE-bench framework use MIT licenses; BFCL uses Apache 2.0.
SWE task content retains the applicable Astropy, Django, or Matplotlib license notices.
AppWorld protected content uses Apache 2.0 with an additional encryption requirement for public redistribution, including derivatives.
We omit protected AppWorld task and solution text and provide metadata and scripts for local reconstruction from official data.

\subsection{Excluded Tasks}
\label{app:v4v-exclusions}
The 45 exclusions comprise five BFCL tasks with conflicts between default-argument rules and answer keys, 35 Terminal-Bench tasks with false passes from the test-output parser, one Terminal-Bench task with browser failures treated as successful checks, one Terminal-Bench task with inconsistent repeated decisions, and three SWE-bench tasks with parsing and empty-report errors.
Table~\ref{vb:excluded-tasks} identifies every excluded task.

\paragraph{Exclusion reasons and supporting evidence.}
\textbf{D: Default-argument conflict.} Public task requirements and the reference answer key disagree about allowed omission of default arguments. Evidence comprises 16 test-case records across four tasks and a comparison of the instruction, schema, answer key, and checker for the fifth.

\noindent\textbf{P: False passes from skipped tests.} The Terminal-Bench pytest parser treats \texttt{SKIPPED}/\texttt{XFAIL} as passing outcomes. Reanalysis of recorded test output provides 248 records covering 190 distinct test cases across all 35 tasks.

\noindent\textbf{B: Browser failures accepted.} Browser startup failure or timeout is treated as absence of an alert, allowing the HTML check to pass; both failure types were examined for \texttt{filter-js-from-html}.

\noindent\textbf{R: Inconsistent decisions.} Repeated valid reference executions return opposite decisions for the same test case; the evidence contains 27 valid runs of \texttt{cross-entropy-method}.

\noindent\textbf{S: Parsing and empty-report errors.} The SWE-bench parser misreads \texttt{<-} in pytest output as a test status, and an empty test-count denominator receives full credit, allowing a no-op test case to pass. Each of the three tasks was checked in three native execution rounds from the correct repository base.

\begin{table*}[!tp]
\centering
\fontsize{10}{12}\selectfont
\setlength{\tabcolsep}{5pt}
\renewcommand{\arraystretch}{1.02}
\begin{tabularx}{\textwidth}{@{}l >{\raggedright\arraybackslash}X c@{}}
\toprule
Source & Task ID & Reason \\
\midrule
BFCL & \nolinkurl{parallel_65} & D \\
SWE & \nolinkurl{astropy__astropy-7671} & S \\
TB & \nolinkurl{feal-differential-cryptanalysis} & P \\
TB & \nolinkurl{catch-me-if-you-can} & P \\
TB & \nolinkurl{amuse-install} & P \\
TB & \nolinkurl{feal-linear-cryptanalysis} & P \\
TB & \nolinkurl{code-from-image} & P \\
BFCL & \nolinkurl{parallel_135} & D \\
TB & \nolinkurl{fix-git} & P \\
TB & \nolinkurl{chess-best-move} & P \\
TB & \nolinkurl{count-call-stack} & P \\
SWE & \nolinkurl{astropy__astropy-7336} & S \\
TB & \nolinkurl{constraints-scheduling} & P \\
TB & \nolinkurl{decommissioning-service-with-sensitive-data} & P \\
TB & \nolinkurl{chem-property-targeting} & P \\
TB & \nolinkurl{dna-insert} & P \\
TB & \nolinkurl{extract-moves-from-video} & P \\
TB & \nolinkurl{countdown-game} & P \\
TB & \nolinkurl{dna-assembly} & P \\
TB & \nolinkurl{db-wal-recovery} & P \\
BFCL & \nolinkurl{parallel_180} & D \\
TB & \nolinkurl{bn-fit-modify} & P \\
TB & \nolinkurl{cross-entropy-method} & R \\
TB & \nolinkurl{assign-seats} & P \\
BFCL & \nolinkurl{parallel_multiple_137} & D \\
BFCL & \nolinkurl{parallel_70} & D \\
TB & \nolinkurl{break-filter-js-from-html} & P \\
TB & \nolinkurl{blind-maze-explorer-5x5} & P \\
TB & \nolinkurl{enemy-grid-escape} & P \\
TB & \nolinkurl{deterministic-tarball} & P \\
TB & \nolinkurl{3d-model-format-legacy} & P \\
SWE & \nolinkurl{astropy__astropy-7166} & S \\
TB & \nolinkurl{cartpole-rl-training} & P \\
TB & \nolinkurl{acl-permissions-inheritance} & P \\
TB & \nolinkurl{configure-git-webserver} & P \\
TB & \nolinkurl{blind-maze-explorer-algorithm} & P \\
TB & \nolinkurl{cobol-modernization} & P \\
TB & \nolinkurl{cancel-async-tasks} & P \\
TB & \nolinkurl{extract-safely} & P \\
TB & \nolinkurl{filter-js-from-html} & B \\
TB & \nolinkurl{distribution-search} & P \\
TB & \nolinkurl{analyze-access-logs} & P \\
TB & \nolinkurl{count-dataset-tokens} & P \\
TB & \nolinkurl{crack-7z-hash} & P \\
TB & \nolinkurl{aimo-airline-departures} & P \\
\bottomrule
\end{tabularx}
\caption{\textbf{All 45 excluded tasks.} BFCL denotes BFCL v4, SWE denotes SWE-bench Verified, and TB denotes Terminal-Bench. Reason codes refer to the exclusion categories described in this section.}
\label{vb:excluded-tasks}
\end{table*}

\subsection{Checks on Retained Tasks}
\label{app:v4v-retained}

The frozen screening records contain one row for each of the 222 retained tasks and mark all six checks as \texttt{pass}, giving 1,332 task--check outcomes.
Table~\ref{vb:qualification-status} summarizes the recorded outcomes and their supporting evidence.

\begin{table*}[!tp]
\centering
\fontsize{10}{12}\selectfont
\setlength{\tabcolsep}{4pt}
\renewcommand{\arraystretch}{1.14}
\begin{tabularx}{\textwidth}{@{}>{\raggedright\arraybackslash}p{1.36in} c >{\raggedright\arraybackslash}X@{}}
\toprule
Check & Recorded passes & Evidence and coverage \\
\midrule
C1: Replay & 222/222 & Reference decisions on all 4,993 cases agree with the labels assigned during construction. \\
C2: Repeatability & 222/222 & Repeated executions under the same initial conditions return stable decisions that agree with the labels. \\
C3: Provenance & 222/222 & All 4,993 case files and 222 task specifications match their frozen snapshots. \\
C4: Input-label consistency & 222/222 & A scan of all 4,993 cases finds no opposite-label groups with identical public inputs within a task. \\
C5: Trivial baselines & 222/222 & Every task has passing records for all three comparisons: always-accept, always-reject, and exact-match. \\
C6: Mutation testing & 222/222 & Each task's suite detects every mutation in its operator set (1,410 configurations total); Section~\ref{app:v4v-mutation} reports the criterion and representative detection evidence. \\
\bottomrule
\end{tabularx}
\caption{\textbf{V4V outcomes and evidence.} The second column reports frozen screening statuses. Checks follow the C1--C6 order in Section~\ref{sec:validation}.}
\label{vb:qualification-status}
\end{table*}

\par\medskip
\noindent\begin{minipage}{\columnwidth}
\paragraph{Mapping to the recorded fields.}
The screening records use the following field names:
\begin{center}
\fontsize{10}{12}\selectfont
\renewcommand{\arraystretch}{1.08}
\begin{tabular}{@{}ll@{}}
\toprule
Check & Recorded field \\
\midrule
C1 & \nolinkurl{C1_replay_recorded} \\
C2 & \nolinkurl{C5a_repeatability_recorded} \\
C3 & \nolinkurl{C4_integrity_recorded} \\
C4 & \nolinkurl{Det_recorded} \\
C5 & \nolinkurl{C2_nontriviality_recorded} \\
C6 & \nolinkurl{C3_mutation_recorded} \\
\bottomrule
\end{tabular}
\end{center}
This mapping separates the paper's numbering from the historical numbering in the source records.
\end{minipage}
\par\medskip

\subsection{Mutation Construction and Detection}
\label{app:v4v-mutation}

Mutation operators are defined manually by removing required checks, replacing conditions with unconditional acceptance, or changing how states and test results are handled.
A task--operator pair defines one configuration, rather than one test case or one execution.
Table~\ref{vb:v4v-mutation-catalogue} lists the 1,410 configurations by source and the checks they modify.
A mutation is detected when the original and modified verifiers both return valid \texttt{PASS}/\texttt{FAIL} decisions on the same test case and those decisions differ.
A crash or invalid output does not count as a detection.

\begin{table*}[!tp]
\centering
\fontsize{10}{12}\selectfont
\setlength{\tabcolsep}{4pt}
\renewcommand{\arraystretch}{1.12}
\begin{tabularx}{\textwidth}{@{}l r r r >{\raggedright\arraybackslash}X@{}}
\toprule
Source & Tasks & Operators/task & Configurations & Checks modified \\
\midrule
BFCL & 45 & 9 & 405 & Required and extra arguments, types, values, function calls, and containers. \\
ACEBench & 50 & 11 & 550 & Total calls, function sets, per-function call counts, and arguments. \\
$\tau^2$-bench & 32 & 3 & 96 & Communication, database, and action reward checks. \\
AgentDojo & 20 & 2 & 40 & Text and final-state checks. \\
AppWorld & 28 & 3 & 84 & Answer comparison, state assertions, and failure handling. \\
SWE-bench & 47 & 5 & 235 & Newly passing and previously passing tests, missing tests, expected failures, and empty reports. \\
\midrule
\textbf{Total} & \textbf{222} & & \textbf{1,410} & \\
\bottomrule
\end{tabularx}
\caption{\textbf{Mutation configurations.} Operators alter reference checks; the counts are task--operator pairs and are not numbers of independent executions.}
\label{vb:v4v-mutation-catalogue}
\end{table*}

A task passes C6 when its test suite detects every mutation in the task's operator set, where detection is defined above; all 222 retained tasks pass, so each mutation across the 1,410 configurations is detected by the corresponding suite.

\paragraph{Representative detection evidence.}
We additionally document detailed evidence for a subset of configurations, drawn from the follow-up records.
For SWE-bench, 94 configurations include a detection example supported by raw test logs matched to the candidate patch contents.
For tool-call tasks, 87 configurations include a detection example from paired mutations.
The paired checks address cases where another check prevents the target mutation from changing the final decision: an additional modification removes that interference so that the target check can be tested.
These two sets illustrate the detection procedure with different evidence formats; they are examples rather than the full set of detected configurations.

For C6, single-target detection requires the original verifier $V$ and its target-mutated version $V_a$ to return different valid \texttt{PASS}/\texttt{FAIL} decisions on the same input under the same execution conditions.
Paired analysis compares $V_b$ with $V_{a,b}$, where the same auxiliary mutation $b$ is applied to both; it provides only conditional detection evidence and is not included in the single-target detection rate.
The 87 paired configurations reported above are not counted as single-target detections.

\subsection{Targeted Rechecks}
\label{app:v4v-audit}

Targeted execution checks provide additional evidence: replay on 16 $\tau^2$ banking tasks covers 320 cases and agrees with their labels; native rechecks of three retained SWE tasks cover 38 cases, each executed three times from the correct repository base, with decisions agreeing with their labels.
These checks support reproducibility for the tested cases, while the 45 exclusions document specific source defects found during screening.

\section{Test Cases and Construction Categories}
\label{app:candidates}

The test suites contain 1,566 positive test cases (31.4\%) and 3,427 negative test cases (68.6\%).
Table~\ref{vb:taxonomy} describes the nine construction categories.
The category records how a test case was constructed and its intended positive or negative outcome; V4V checks the assigned label against the source verifier before the test suite is fixed.
Neither category names nor labels are supplied during generation.
In the main text, as grouped in Table~\ref{vb:composition}, \texttt{cosmetic} and \texttt{overstrict\_trap} cases together form the valid variants; \texttt{base}, \texttt{partial}, \texttt{near\_miss}, and \texttt{regression} form the ordinary errors; and \texttt{hack} and \texttt{process\_violation} form the attacks.
\begin{table*}[!tp]
\centering
\fontsize{10}{12}\selectfont
\setlength{\tabcolsep}{4pt}
\renewcommand{\arraystretch}{1.15}

\begin{tabularx}{\textwidth}{@{}l l r >{\raggedright\arraybackslash}X@{}}
\toprule
Label & Class & Count & Construction and intended distinction \\
\midrule
\texttt{PASS} & \texttt{gold} & 222 & Reference solution; tests acceptance of the designated reference solution. \\
\texttt{PASS} & \texttt{cosmetic} & 840 & Changes formatting or other irrelevant representation details; tests acceptance of equivalent presentations. \\
\texttt{PASS} & \texttt{overstrict\_trap} & 504 & Uses an alternative implementation or execution path; tests acceptance of valid alternatives. \\
\midrule
\texttt{FAIL} & \texttt{base} & 222 & Uses an initial, unrepaired, or otherwise unfinished outcome; tests recognition of basic task failure. \\
\texttt{FAIL} & \texttt{partial} & 999 & Removes required arguments, actions, or changes; tests whether all required parts are checked. \\
\texttt{FAIL} & \texttt{near\_miss} & 907 & Alters a consequential value or local behavior; tests detection of subtle errors. \\
\texttt{FAIL} & \texttt{regression} & 501 & Introduces an additional failure or side effect; tests preservation of other requirements. \\
\texttt{FAIL} & \texttt{hack} & 796 & Exploits checking omissions, fabricates success signals, or interferes with verification; tests resistance to these attacks. \\
\texttt{FAIL} & \texttt{process\_violation} & 2 & Violates a safety constraint; both current examples come from AgentDojo injection tasks. \\
\midrule
\multicolumn{2}{@{}l}{\textbf{Total}} & \textbf{4993} & 1,566 \texttt{PASS} and 3,427 \texttt{FAIL}. \\
\bottomrule
\end{tabularx}
\caption{\textbf{Test-case categories.} Construction and reference labels for the final 4,993 test cases. All final leaderboard configurations use the same task set. Test cases are withheld during verifier generation.}
\label{vb:taxonomy}
\end{table*}

For the six-group diagnostics, ordinary errors combine \texttt{base}, \texttt{partial}, \texttt{near\_miss}, and \texttt{regression} (2,629 test cases).
The remaining groups are reference solutions (222), cosmetic variants (840), alternative implementations or paths (504), hacks (796), and process violations (2).
The last two groups form the 798 attack test cases.

\paragraph{Tool-call test cases.}
For these test cases, the verifier receives raw call text and a parsed representation of function names and arguments.
The task instruction specifies the requested operations, and the environment provides the tool definitions.
For the DNA-analysis task in Appendix~\ref{app:cases}, those definitions state that \texttt{mutation\_type} defaults to \texttt{insertion}; omitting that optional argument therefore preserves the requested operation.
This explicit rule provides a basis for identifying the generated verifier's incorrect treatment of the input.

\paragraph{Code test cases.}
In task \texttt{django-11740}, the instruction describes missing migration dependencies, and the environment provides a repository checkout.
At evaluation time, the generated verifier receives a candidate patch through a \texttt{diff} field.
The recorded generation process inspects the migration autodetector and runs the repository's tests through the public execution tool.
It then constructs behavior checks for the proposed patch and checks existing migration behavior.
This instance demonstrates access to source code and executed tests, rather than reliance only on a candidate-authored success report.
It does not independently certify the retained reference or the integrity of every code-task environment.

\section{Metrics and Scoring Protocol}
\label{app:metrics}

\paragraph{Outcome categories.}
For each test case, $P$ denotes a valid \texttt{PASS}, $F$ a valid \texttt{FAIL}, and $U$ no valid decision.
The last category includes missing verifiers and invocations that fail to return a conforming decision.
Positive $F$ outcomes are false rejections, while negative $P$ outcomes are false acceptances.
Both differ from $U$, even though all receive zero correctness credit on the affected test case.
Optional confidence values do not change the scoring label: a well-formed \texttt{FAIL} with low or missing confidence remains $F$.

\paragraph{Equal task weighting.}
Let $\mathcal C_t^+$ and $\mathcal C_t^-$ be the positive and negative test cases for task $t$.
If $A_{m,t}$ counts positive test cases receiving \texttt{PASS} and $R_{m,t}$ counts negative test cases receiving \texttt{FAIL}, define
\begin{align}
 r^+_{m,t}&=\frac{A_{m,t}}{|\mathcal C_t^+|},\\
 r^-_{m,t}&=\frac{R_{m,t}}{|\mathcal C_t^-|}.
\end{align}
Both denominators include test cases with missing or invalid outputs.
All tasks have positive and negative test cases, so both denominators are nonzero.
For the full task set $\mathcal D$, scores in percentage units are
\begin{align}
 \mathrm{TPR}_m&=\frac{100}{|\mathcal D|}\sum_{t\in\mathcal D}r^+_{m,t},\\
 \mathrm{TNR}_m&=\frac{100}{|\mathcal D|}\sum_{t\in\mathcal D}r^-_{m,t},\\
 \mathrm{BA}_m&=\frac{\mathrm{TPR}_m+\mathrm{TNR}_m}{2},\\
 \mathrm{BA}_{\rm norm,m}&=2\mathrm{BA}_m-100.
\end{align}
A missing verifier contributes zero to both class-specific rates for its task.
Always accepting or always rejecting yields BA$_{\rm norm}=0$ when a valid decision is returned throughout; independent random binary decisions have the same expected score.
Perfect agreement yields 100, and no valid decisions on any test case yields $-100$.
Normalization preserves the ordering induced by task-averaged BA.

\paragraph{Attack acceptance.}
Let $\mathcal H_t$ contain the \texttt{hack} and \texttt{process\_violation} test cases for task $t$.
Each of the 222 retained tasks has at least one such test case.
If $B_{m,t}$ counts attack test cases receiving a valid \texttt{PASS}, the attack false-positive rate is
\begin{equation}
 \mathrm{FPR}_{h,m}=\frac{100}{|\mathcal D|}
 \sum_{t\in\mathcal D}
 \frac{B_{m,t}}{|\mathcal H_t|}.
\end{equation}
An invalid or missing output does not count as an accepted attack.
Consequently, low FPR$_h$ must be interpreted alongside valid-output coverage and the primary accuracy metric.

\paragraph{Pooled diagnostics.}
Submission is the number of tasks with a submitted verifier divided by 222.
Valid-output rate is $100(P+F)/(P+F+U)$ over the relevant test cases.
On positive test cases, rejection share is $100F/(F+U)$, the fraction of unaccepted test cases receiving an explicit rejection.
These diagnostic proportions weight test cases equally, whereas the main metrics weight tasks equally.

\paragraph{Harness comparisons and rounding.}
Every final harness pair uses the same 222 tasks, with missing results retained in the denominator.
For configurations $a$ and $b$, $\Delta S=S_a(\mathcal D)-S_b(\mathcal D)$.
Scores and source-reported differences are rounded independently to one decimal place, which can produce a 0.1-point discrepancy when subtracting displayed values.
The reported ranking uses point estimates rather than significance tests or repeated-generation estimates.

\section{Generation Configurations}
\label{app:setup}

The Pi harness comparison contains 19 models.
GPT-5.6-sol, Qwen3.7 Plus, and GPT-6-astra additionally use OpenClaw, mini-swe, Codex, and Claude Code.
GPT-5.6-terra and GPT-5.5 additionally use Codex; two Gemini models use Gemini CLI; and Claude Opus 4.8 uses Claude Code.
This gives 17 alternative configurations and 36 distinct configurations overall.
Table~\ref{vb:harness} reports the within-model comparisons, and Table~\ref{vb:alt-results} gives full metrics for the 17 alternative configurations.
\begin{table*}[!tp]
\centering

\fontsize{10}{12}\selectfont
\setlength{\tabcolsep}{2.5pt}
\renewcommand{\arraystretch}{1.10}
\begin{tabular*}{\linewidth}{@{\extracolsep{\fill}}llrrrrr@{}}
\toprule
Model & Alternative & \shortstack{BA$_{\rm norm}$\\ Pi harness} & \shortstack{BA$_{\rm norm}$\\ Alternative} & $\Delta$BA$_{\rm norm}$ & \shortstack{Submitted/$N$\\ Pi harness} & \shortstack{Submitted/$N$\\ Alternative} \\
\midrule
Claude Opus 4.8 & Claude Code & 54.0 & 66.2 & $+12.2$ & 218/222 & 220/222 \\
GPT-6-astra & Claude Code & 62.1 & 61.1 & $-0.9$ & 222/222 & 222/222 \\
GPT-6-astra & Codex & 62.1 & 58.0 & $-4.1$ & 222/222 & 219/222 \\
GPT-6-astra & mini-swe & 62.1 & 54.9 & $-7.2$ & 222/222 & 222/222 \\
GPT-6-astra & OpenClaw & 62.1 & 51.1 & $-10.9$ & 222/222 & 216/222 \\
Gemini 3.5 Flash & Gemini CLI & 35.7 & 50.7 & $+15.0$ & 222/222 & 221/222 \\
Qwen3.7 Plus & Claude Code & 40.2 & 48.9 & $+8.7$ & 222/222 & 222/222 \\
GPT-5.6-sol & Claude Code & 52.9 & 46.3 & $-6.5$ & 219/222 & 220/222 \\
Qwen3.7 Plus & mini-swe & 40.2 & 40.5 & $+0.3$ & 222/222 & 221/222 \\
Qwen3.7 Plus & OpenClaw & 40.2 & 39.9 & $-0.3$ & 222/222 & 219/222 \\
GPT-5.6-sol & mini-swe & 52.9 & 35.8 & $-17.1$ & 219/222 & 218/222 \\
GPT-5.6-sol & OpenClaw & 52.9 & 35.1 & $-17.8$ & 219/222 & 216/222 \\
GPT-5.6-terra & Codex & 52.9 & 31.0 & $-21.8$ & 221/222 & 214/222 \\
GPT-5.5 & Codex & 46.6 & 30.3 & $-16.4$ & 218/222 & 219/222 \\
GPT-5.6-sol & Codex & 52.9 & 26.2 & $-26.6$ & 219/222 & 218/222 \\
Qwen3.7 Plus & Codex & 40.2 & 26.1 & $-14.1$ & 222/222 & 222/222 \\
Gemini 2.5 Pro & Gemini CLI & 14.2 & 5.0 & $-9.2$ & 220/222 & 216/222 \\
\bottomrule
\end{tabular*}
\par\smallskip
\begin{minipage}{\linewidth}\footnotesize
The Pi harness columns reproduce Table~\ref{vb:main-results}. All rows and paired differences use the same 222 tasks. Source-reported differences are rounded independently from absolute scores.
\end{minipage}
\caption{\textbf{Harness comparison.} Configurations ordered by the alternative configuration's normalized BA. $\Delta$BA$_{\rm norm}$ is the alternative harness score minus the Pi harness score, in normalized-BA points. Full metrics appear in Table~\ref{vb:main-results} for the Pi harness and in Table~\ref{vb:alt-results} for the alternatives.}
\label{vb:harness}
\end{table*}

\paragraph{Changes in acceptance and rejection.}
For Qwen3.7 Plus, Codex raises TNR from 82.6\% with the Pi harness to 92.1\%, but lowers TPR from 57.6\% to 34.0\%; BA$_{\rm norm}$ falls from 40.2 to 26.1.
Claude Opus 4.8 with Claude Code improves both components: TPR rises from 67.7\% to 75.8\% and TNR from 86.3\% to 90.4\%, increasing BA$_{\rm norm}$ by 12.2 points.
Thus, changes across configurations cannot be reduced to a necessary trade-off between acceptance and rejection.
These comparisons do not isolate the effect of a single harness component.

\begin{table*}[!tp]
\centering
\fontsize{10}{12}\selectfont
\setlength{\tabcolsep}{3pt}
\renewcommand{\arraystretch}{1.08}
\begin{tabular*}{\linewidth}{@{\extracolsep{\fill}}llrrrr@{}}
\toprule
Model & Harness & BA$_{\rm norm}$ & TPR & TNR & FPR$_h$ \\
\midrule
Claude Opus 4.8 & Claude Code & \textbf{66.2} & \textbf{75.8} & 90.4 & 7.8 \\
GPT-6-astra & Claude Code & 61.1 & 68.0 & 93.2 & 9.6 \\
GPT-6-astra & Codex & 58.0 & 66.2 & 91.8 & 9.2 \\
GPT-6-astra & mini-swe & 54.9 & 61.9 & 93.0 & 9.5 \\
GPT-6-astra & OpenClaw & 51.1 & 63.2 & 87.9 & 9.1 \\
Gemini 3.5 Flash & Gemini CLI & 50.7 & 59.2 & 91.5 & 7.7 \\
Qwen3.7 Plus & Claude Code & 48.9 & 59.7 & 89.1 & 11.5 \\
GPT-5.6-sol & Claude Code & 46.3 & 49.0 & \textbf{97.4} & 2.3 \\
Qwen3.7 Plus & mini-swe & 40.5 & 53.3 & 87.1 & 12.7 \\
Qwen3.7 Plus & OpenClaw & 39.9 & 51.2 & 88.6 & 10.7 \\
GPT-5.6-sol & mini-swe & 35.8 & 38.4 & \textbf{97.4} & 1.1 \\
GPT-5.6-sol & OpenClaw & 35.1 & 38.7 & 96.4 & \textbf{0.6} \\
GPT-5.6-terra & Codex & 31.0 & 39.2 & 91.8 & 4.7 \\
GPT-5.5 & Codex & 30.3 & 41.2 & 89.1 & 6.2 \\
GPT-5.6-sol & Codex & 26.2 & 32.3 & 93.9 & 1.0 \\
Qwen3.7 Plus & Codex & 26.1 & 34.0 & 92.1 & 6.5 \\
Gemini 2.5 Pro & Gemini CLI & 5.0 & 23.7 & 81.3 & 3.5 \\
\bottomrule
\end{tabular*}
\par\smallskip
\begin{minipage}{\linewidth}\footnotesize
BA$_{\rm norm}=\mathrm{TPR}+\mathrm{TNR}-100$; TPR, TNR, and FPR$_h$ are percentages. Scores are rounded independently to one decimal place. Bold marks the best BA$_{\rm norm}$, TPR, and TNR and the lowest FPR$_h$ among these 17 configurations.
\end{minipage}
\caption{\textbf{Full results for alternative harness configurations.} The 17 configurations outside the common Pi harness, ordered by BA$_{\rm norm}$ and scored on the same 222 tasks as Table~\ref{vb:main-results}. Differences across harnesses describe configurations rather than isolated model capability.}
\label{vb:alt-results}
\end{table*}

\paragraph{Budgets and tools.}
The global limits are 5,400 seconds per task and 32,768 output tokens per reply.
There is no additional benchmark-wide cap on total output tokens per task or context length, although model and harness limits apply.
Generation tools permit task inspection, public environment execution, and reset where supported, together with the harness's code-authoring tools.
Test cases and reference implementations are withheld during generation.

\paragraph{Harness and reasoning settings.}
Recorded versions include Pi harness 0.84.2, Codex 0.149.0, Claude Code 2.1.263, Gemini CLI 0.60.0, OpenClaw 2026.6.34, and mini-swe-agent 2.4.6.
Reasoning settings also differ.
For example, retained structured Opus/Claude Code traces record high effort with thinking enabled, while historical Pi harness traces that expose a thinking setting record it as off.
Later configurations include other settings.

\paragraph{Generation instructions.}
The verifier-generation prompts require a binary \texttt{PASS}/\texttt{FAIL} decision and provide optional score or confidence fields that do not affect the binary outcome used for scoring.

\paragraph{Retained attempts.}
The release selects the latest authorized attempt for a task and configuration, including attempts that do not deliver a verifier; it does not select the highest-scoring attempt.
Earlier attempts remain in the attempt manifest and are not combined with a later verifier's judgments in the case analysis.

\paragraph{Judgment rounds.}
Tool-call tasks use the first judgment round, while state and code tasks are judged three times.
For each aggregate round, metrics are computed over the full 222-task pool with the first-round tool-call judgments held fixed.
The round with median aggregate BA determines the entire reported row; columns are not selected independently.
The grouped diagnostics use the same selected verifiers and judgment rounds as the main table.
These rounds execute fixed generated verifiers and do not constitute independent generations.
Native traces retain task-specific prompts and available configuration records, although complete system prompts and structured traces are not uniformly available.

\paragraph{Reasoning-effort ablation.}
Table~\ref{vb:reasoning} reports the Astra trace batch introduced in Section~\ref{sec:setup}, a separate generation of GPT-6-astra verifiers under LOW and HIGH reasoning effort with the Pi harness and Claude Code.
Each condition uses all 222 tasks, a 5,400-second limit per task, and at most 32,768 output tokens per response.
For tasks supporting environment execution, the exploration budget allows 20 \texttt{run\_public} executions per task, shared by direct tool calls and calls through \texttt{vgb\_env}.
This limit does not count all tool calls or conversation turns.
All four conditions deliver verifiers for all 222 tasks.
Scoring follows the same full-task and median-round protocol as the main evaluation.
The ablation conditions are not substituted for the main-table Pi harness or Claude Code configurations.

\begin{table*}[!tp]
\centering
\fontsize{10}{12}\selectfont
\setlength{\tabcolsep}{5pt}
\begin{tabular*}{\linewidth}{@{\extracolsep{\fill}}llrrrrr@{}}
\toprule
Harness & Effort & BA$_{\rm norm}$ & TPR & TNR & FPR$_h$ & Submitted/$N$ \\
\midrule
Pi harness & LOW & 59.0 & 64.0 & 95.0 & 7.3 & 222/222 \\
Pi harness & HIGH & 59.5 & 66.8 & 92.7 & 9.5 & 222/222 \\
 & $\Delta$ HIGH--LOW & $+0.5$ & $+2.8$ & $-2.3$ & $+2.2$ & 0 \\
\midrule
Claude Code & LOW & 60.5 & 67.2 & 93.3 & 9.4 & 222/222 \\
Claude Code & HIGH & 59.2 & 66.6 & 92.6 & 9.7 & 222/222 \\
 & $\Delta$ HIGH--LOW & $-1.3$ & $-0.6$ & $-0.7$ & $+0.4$ & 0 \\
\bottomrule
\end{tabular*}
\caption{\textbf{Reasoning-effort ablation for GPT-6-astra.} Each condition uses the same 222 tasks and stated generation budgets. Differences are computed from unrounded scores and may differ from subtraction of displayed values. These four conditions form the Astra trace batch, which is separate from the main release in Table~\ref{vb:main-results}.}
\label{vb:reasoning}

\par\vspace{12pt}
\begin{minipage}[t]{0.48\textwidth}
\vspace{0pt}
\centering
\fontsize{10}{12}\selectfont
\setlength{\tabcolsep}{3pt}
\begin{tabular*}{\linewidth}{@{\extracolsep{\fill}}lrrr@{}}
\toprule
Task type & $N$ & Pi harness & Claude Code \\
\midrule
Tool-call & 95 & $+3.9$ & $-0.6$ \\
State & 80 & $-5.4$ & $-3.6$ \\
Code & 47 & $+3.8$ & $+1.4$ \\
\bottomrule
\end{tabular*}
\caption{\textbf{Reasoning-effort differences by task type.} HIGH minus LOW, in BA$_{\rm norm}$ points, in the Astra trace batch. $N$ is the number of tasks in each type.}
\label{vb:effort-types}

\end{minipage}\hfill
\begin{minipage}[t]{0.48\textwidth}
\vspace{0pt}
\centering
\fontsize{10}{12}\selectfont
\setlength{\tabcolsep}{3pt}
\begin{tabular*}{\linewidth}{@{\extracolsep{\fill}}llrr@{}}
\toprule
Harness & Effort & \shortstack{State/code\\blocked} & \shortstack{$\tau^2$ before\\first write} \\
\midrule
Pi harness & LOW & 4/127 & 0/32 \\
Pi harness & HIGH & 77/127 & 30/32 \\
Claude Code & LOW & 21/127 & 7/32 \\
Claude Code & HIGH & 98/127 & 20/32 \\
\bottomrule
\end{tabular*}
\caption{\textbf{Execution-budget exhaustion during generation.} A task is counted once if any request exceeds the generation budget. The final column counts $\tau^2$-bench tasks whose first blocked request precedes the first write of the verifier. These are generation-stage observations, not runtime failures during hidden evaluation.}
\label{vb:effort-budget}

\end{minipage}

\end{table*}

\paragraph{Variation across task types.}
Table~\ref{vb:effort-types} decomposes the reasoning-effort comparison by task type, using the same selected verifiers and judgment rounds as Table~\ref{vb:reasoning}.
Within each type, BA$_{\rm norm}$ gives tasks equal weight.
The changes have different signs across types: HIGH improves code-task scores with both harnesses but reduces state-task scores.
In the Pi harness tool-call subset, HIGH additionally accepts 63 positive test cases, with no positive test case changing from accepted to rejected; it also introduces 128 false acceptances and corrects three.
Thus, the increase in positive acceptance does not uniformly improve discrimination between positive and negative test cases.

\paragraph{Execution budget during generation.}
Table~\ref{vb:effort-budget} counts final selected generation attempts in which at least one environment-execution request exceeded the exploration budget.
We use the generation record's \texttt{cap\_budget\_blocks} counter; the judgment-stage counters are separate.
The denominator contains the 127 state and code tasks, which support environment execution.
The \texttt{cap\_run\_public} counter includes blocked requests and is therefore not a count of successful executions.

To locate budget exhaustion within generation, we compare the first native tool result containing \nolinkurl{BudgetExceeded: run_public} with the first explicit write to \texttt{src/main.py}.
All 127 selected state and code traces per condition contain such a write.
The final column reports this ordering on the 32 $\tau^2$-bench tasks.
It identifies budget exhaustion before the verifier is first written, without treating tool counts as a complete annotation of self-testing.

\section{Results by Task Type}
\label{app:task-types}

Tables~\ref{vb:tool-ranking}--\ref{vb:code-ranking} disaggregate the 19 Pi harness configurations in Table~\ref{vb:main-results} into 95 tool-call, 80 state, and 47 code tasks.
They use the same generated verifiers and judgment rounds as the overall ranking.
Each task receives equal weight within its type, and missing or invalid outputs remain in the denominator.
Ranks follow unrounded $\mathrm{BA}_{\mathrm{norm}}$; displayed values are rounded to one decimal place.
The additional harness configurations are not included in this breakdown.\looseness=-1

\begin{table*}[!tp]
\centering
\fontsize{10}{12}\selectfont
\setlength{\tabcolsep}{5pt}
\renewcommand{\arraystretch}{1.05}
\begin{tabular*}{\textwidth}{@{\extracolsep{\fill}}rlrrrr@{}}
\toprule
Rank & Model & $\mathrm{BA}_{\mathrm{norm}}\uparrow$ & TPR$\uparrow$ & TNR$\uparrow$ & $\mathrm{FPR}_{h}\downarrow$ \\
\midrule
1 & GPT-5.6-sol & \textbf{85.2} & 86.5 & 98.7 & 1.8 \\
2 & GPT-6-astra & 83.0 & 92.8 & 90.2 & 17.1 \\
3 & Kimi K2.6 & 79.6 & 87.1 & 92.5 & 10.4 \\
4 & GLM-5.2 & 75.9 & 93.3 & 82.6 & 25.7 \\
5 & DeepSeek V4 Flash & 74.7 & 87.8 & 86.9 & 18.2 \\
6 & GPT-5.6-terra & 74.6 & 77.6 & 96.9 & 6.1 \\
7 & Kimi K3 & 72.8 & 92.2 & 80.6 & 31.0 \\
8 & DeepSeek V4 Pro & 71.0 & 88.3 & 82.7 & 22.7 \\
9 & Claude Opus 4.8 & 68.0 & 92.1 & 75.9 & 31.1 \\
10 & Doubao Seed 2.1 Pro & 67.3 & 83.2 & 84.1 & 22.5 \\
11 & GPT-5.5 & 65.6 & 87.4 & 78.2 & 23.2 \\
12 & Gemini 3.5 Flash & 64.6 & 87.9 & 76.7 & 30.5 \\
13 & GPT-5.1 & 59.1 & 73.9 & 85.2 & 20.4 \\
14 & Qwen3.7 Plus & 57.5 & 73.7 & 83.9 & 21.9 \\
15 & Doubao Seed 2.0 Code & 56.2 & 73.7 & 82.5 & 23.3 \\
16 & Gemini 2.5 Pro & 54.8 & 62.8 & 92.0 & 8.1 \\
17 & Doubao Seed 2.0 Lite & 53.7 & 70.9 & 82.9 & 20.4 \\
18 & Doubao Seed 2.0 Pro & 44.3 & 52.2 & 92.1 & 13.8 \\
19 & Doubao Seed 2.0 Mini & 26.9 & 42.2 & 84.7 & 7.8 \\
\bottomrule
\end{tabular*}
\caption{\textbf{Tool-call verification ranking.} All 19 models use the Pi harness, evaluated on the same 95 tasks. Rates are percentages; $\mathrm{BA}_{\mathrm{norm}}$ is reported in points. Ranks use unrounded scores.}
\label{vb:tool-ranking}
\end{table*}

\begin{table*}[!tp]
\centering
\fontsize{10}{12}\selectfont
\setlength{\tabcolsep}{5pt}
\renewcommand{\arraystretch}{1.05}
\begin{tabular*}{\textwidth}{@{\extracolsep{\fill}}rlrrrr@{}}
\toprule
Rank & Model & $\mathrm{BA}_{\mathrm{norm}}\uparrow$ & TPR$\uparrow$ & TNR$\uparrow$ & $\mathrm{FPR}_{h}\downarrow$ \\
\midrule
1 & Claude Opus 4.8 & \textbf{34.6} & 41.2 & 93.3 & 0.5 \\
2 & Qwen3.7 Plus & 30.7 & 49.0 & 81.7 & 10.5 \\
3 & Doubao Seed 2.1 Pro & 30.0 & 36.1 & 93.9 & 3.6 \\
4 & GPT-5.6-terra & 30.0 & 35.0 & 95.0 & 2.8 \\
5 & GPT-6-astra & 29.1 & 31.0 & 98.1 & 0.2 \\
6 & Gemini 3.5 Flash & 27.6 & 33.6 & 94.0 & 0.5 \\
7 & GPT-5.5 & 19.0 & 32.3 & 86.7 & 0.0 \\
8 & Kimi K3 & 16.5 & 38.6 & 77.9 & 0.5 \\
9 & GPT-5.6-sol & 16.1 & 23.0 & 93.1 & 0.0 \\
10 & Doubao Seed 2.0 Code & 13.9 & 27.5 & 86.4 & 6.5 \\
11 & GLM-5.2 & 11.3 & 39.9 & 71.4 & 3.0 \\
12 & Doubao Seed 2.0 Lite & 9.2 & 21.1 & 88.1 & 5.0 \\
13 & Doubao Seed 2.0 Pro & 5.0 & 7.5 & 97.5 & 2.5 \\
14 & Doubao Seed 2.0 Mini & 0.8 & 4.7 & 96.1 & 0.9 \\
15 & DeepSeek V4 Pro & 0.6 & 33.8 & 66.9 & 1.1 \\
16 & GPT-5.1 & -4.8 & 9.8 & 85.4 & 2.8 \\
17 & Kimi K2.6 & -5.6 & 23.6 & 70.8 & 0.5 \\
18 & DeepSeek V4 Flash & -7.8 & 28.3 & 63.8 & 0.5 \\
19 & Gemini 2.5 Pro & -10.6 & 10.9 & 78.5 & 1.5 \\
\bottomrule
\end{tabular*}
\caption{\textbf{State-based verification ranking.} All 19 models use the Pi harness, evaluated on the same 80 tasks. Rates are percentages; $\mathrm{BA}_{\mathrm{norm}}$ is reported in points. Ranks use unrounded scores.}
\label{vb:state-ranking}
\end{table*}

\begin{table*}[!tp]
\centering
\fontsize{10}{12}\selectfont
\setlength{\tabcolsep}{5pt}
\renewcommand{\arraystretch}{1.05}
\begin{tabular*}{\textwidth}{@{\extracolsep{\fill}}rlrrrr@{}}
\toprule
Rank & Model & $\mathrm{BA}_{\mathrm{norm}}\uparrow$ & TPR$\uparrow$ & TNR$\uparrow$ & $\mathrm{FPR}_{h}\downarrow$ \\
\midrule
1 & Kimi K3 & \textbf{76.9} & 79.5 & 97.5 & 0.0 \\
2 & GPT-6-astra & 75.8 & 78.3 & 97.5 & 0.0 \\
3 & DeepSeek V4 Pro & 66.8 & 73.6 & 93.2 & 0.0 \\
4 & Claude Opus 4.8 & 59.0 & 63.4 & 95.6 & 0.0 \\
5 & GLM-5.2 & 58.1 & 65.1 & 93.0 & 0.0 \\
6 & GPT-5.5 & 55.4 & 59.1 & 96.2 & 0.0 \\
7 & DeepSeek V4 Flash & 53.9 & 61.3 & 92.7 & 0.0 \\
8 & GPT-5.6-sol & 50.1 & 54.5 & 95.6 & 0.0 \\
9 & GPT-5.6-terra & 47.9 & 50.6 & 97.3 & 0.0 \\
10 & Doubao Seed 2.1 Pro & 34.2 & 39.5 & 94.8 & 2.1 \\
11 & Qwen3.7 Plus & 21.4 & 40.0 & 81.4 & 2.1 \\
12 & Kimi K2.6 & 15.0 & 31.9 & 83.1 & 0.0 \\
13 & Doubao Seed 2.0 Pro & 9.9 & 11.2 & 98.7 & 0.0 \\
14 & Doubao Seed 2.0 Code & 3.1 & 16.6 & 86.5 & 0.0 \\
15 & Doubao Seed 2.0 Lite & 2.3 & 14.9 & 87.4 & 0.0 \\
16 & Doubao Seed 2.0 Mini & -5.6 & 6.4 & 88.0 & 0.0 \\
17 & GPT-5.1 & -6.4 & 0.0 & 93.6 & 0.0 \\
18 & Gemini 3.5 Flash & -8.8 & 12.8 & 78.5 & 0.0 \\
19 & Gemini 2.5 Pro & -25.9 & 4.3 & 69.9 & 0.0 \\
\bottomrule
\end{tabular*}
\caption{\textbf{Code-based verification ranking.} All 19 models use the Pi harness, evaluated on the same 47 tasks. Rates are percentages; $\mathrm{BA}_{\mathrm{norm}}$ is reported in points. Ranks use unrounded scores.}
\label{vb:code-ranking}
\end{table*}

\paragraph{Model rankings depend on task type.}
GPT-5.6-sol leads tool-call verification at 85.2, Claude Opus 4.8 leads state-based verification at 34.6, and Kimi K3 leads code-based verification at 76.9.
Ranks also reverse: Kimi K2.6 ranks third on tool-call tasks but seventeenth on state tasks, whereas Qwen3.7 Plus ranks fourteenth and second, respectively.
DeepSeek V4 Pro ranks third on code tasks and fifteenth on state tasks.
These comparisons concern the benchmark's task groups, which also differ in source benchmarks, environments, and test-case composition; they do not isolate the effect of task type alone.\looseness=-1

\paragraph{The acceptance deficit is strongest on state and code tasks.}
Across the 19 models, mean TPR is 79.2\% on tool-call tasks, 27.7\% on state tasks, and 40.2\% on code tasks, while mean TNR is 85.8\%, 85.2\%, and 90.6\%, respectively.
These are arithmetic means of the models' task-averaged rates.
All 19 models have lower TPR than TNR on state and code tasks, but eight have higher TPR than TNR on tool-call tasks.
The aggregate asymmetry therefore coexists with a different pattern on some task types.
State-task TPR remains below 50\% for every model, even for the highest-ranked configuration.
Low positive acceptance includes both explicit rejection and missing or invalid decisions; the rates alone cannot distinguish their causes.\looseness=-1

\paragraph{Low attack acceptance does not establish reliable verification.}
On state tasks, GPT-5.6-sol has $\mathrm{FPR}_{h}=0.0\%$ but accepts only 23.0\% of positives, yielding $\mathrm{BA}_{\mathrm{norm}}=16.1$.
On code tasks, 17 models have $\mathrm{FPR}_{h}=0.0\%$ at the reported precision, yet their $\mathrm{BA}_{\mathrm{norm}}$ scores range from $-25.9$ to 76.9.
A low attack-acceptance rate must therefore be interpreted alongside acceptance of valid solutions and successful rejection of negatives.
Appendix~\ref{app:cross-model-cases} compares input handling, checking rules, and execution on identical test cases.\looseness=-1

\section{Grouped Test-Case Results}
\label{app:extended}

Tables~\ref{vb:group-counts} and~\ref{vb:negative-group-counts} give the supplied final diagnostic counts for all 19 Pi harness models.
Every entry has $P+F+U=N$.
Every model has all 4,993 test cases, including 1,566 positives and 3,427 negatives.
Routing failures and missing submissions remain in these denominators.
These counts use the same verifier selection and judgment rounds as the final leaderboard; the grouped proportions weight test cases, whereas the leaderboard gives tasks equal weight.
\begin{table*}[!tp]
\centering
\fontsize{10}{12}\selectfont
\setlength{\tabcolsep}{5pt}
\renewcommand{\arraystretch}{1.10}
\begin{tabular*}{\textwidth}{@{\extracolsep{\fill}}lrrr@{}}
\toprule
\multicolumn{4}{c}{\textbf{(a) Positive test cases}}\\
\midrule
Model & Gold & Cosmetic & Valid alternative \\
\midrule
GPT-6-astra & 154/67/1 & 573/263/4 & 285/215/4 \\
Claude Opus 4.8 & 152/65/5 & 578/243/19 & 323/172/9 \\
Kimi K3 & 156/49/17 & 590/186/64 & 323/119/62 \\
GPT-5.6-sol & 131/85/6 & 498/319/23 & 257/231/16 \\
GPT-5.6-terra & 131/89/2 & 500/332/8 & 240/259/5 \\
GLM-5.2 & 154/46/22 & 585/169/86 & 316/121/67 \\
Doubao Seed 2.1 Pro & 129/89/4 & 495/332/13 & 283/215/6 \\
GPT-5.5 & 139/65/18 & 528/241/71 & 279/168/57 \\
DeepSeek V4 Pro & 148/44/30 & 556/166/118 & 293/115/96 \\
DeepSeek V4 Flash & 137/54/31 & 519/200/121 & 275/132/97 \\
Qwen3.7 Plus & 131/83/8 & 506/307/27 & 284/211/9 \\
Gemini 3.5 Flash & 118/86/18 & 466/313/61 & 292/190/22 \\
Kimi K2.6 & 120/70/32 & 465/254/121 & 247/170/87 \\
Doubao Seed 2.0 Code & 104/107/11 & 408/394/38 & 239/249/16 \\
Doubao Seed 2.0 Lite & 95/116/11 & 373/426/41 & 216/271/17 \\
Doubao Seed 2.0 Pro & 63/159/0 & 248/592/0 & 142/362/0 \\
GPT-5.1 & 84/125/13 & 336/455/49 & 185/291/28 \\
Gemini 2.5 Pro & 73/114/35 & 290/424/126 & 163/274/67 \\
Doubao Seed 2.0 Mini & 51/154/17 & 200/576/64 & 102/371/31 \\
\bottomrule
\end{tabular*}
\caption{\textbf{Decisions on positive test cases.} Grouped counts on positive test cases for all 19 configurations using the Pi harness. Each cell is $P/F/U$: valid \texttt{PASS}, valid \texttt{FAIL}, and no valid decision. The cell denominator is their sum, $N=P+F+U$.}
\label{vb:group-counts}
\end{table*}
\begin{table*}[!tp]
\centering
\fontsize{10}{12}\selectfont
\setlength{\tabcolsep}{5pt}
\renewcommand{\arraystretch}{1.10}
\begin{tabular*}{\textwidth}{@{\extracolsep{\fill}}lrrr@{}}
\toprule
\multicolumn{4}{c}{\textbf{(b) Negative test cases}}\\
\midrule
Model & Ordinary errors & Hack & Process violation \\
\midrule
GPT-6-astra & 144/2477/8 & 83/711/2 & 1/1/0 \\
Claude Opus 4.8 & 390/2197/42 & 156/628/12 & 2/0/0 \\
Kimi K3 & 311/2220/98 & 156/593/47 & 2/0/0 \\
GPT-5.6-sol & 35/2548/46 & 9/775/12 & 0/2/0 \\
GPT-5.6-terra & 64/2541/24 & 38/754/4 & 1/1/0 \\
GLM-5.2 & 294/2176/159 & 135/599/62 & 2/0/0 \\
Doubao Seed 2.1 Pro & 265/2324/40 & 125/660/11 & 2/0/0 \\
GPT-5.5 & 226/2175/228 & 116/615/65 & 0/2/0 \\
DeepSeek V4 Pro & 232/2191/206 & 117/585/94 & 2/0/0 \\
DeepSeek V4 Flash & 191/2217/221 & 90/616/90 & 2/0/0 \\
Qwen3.7 Plus & 378/2170/81 & 130/644/22 & 2/0/0 \\
Gemini 3.5 Flash & 329/2153/147 & 157/601/38 & 2/0/0 \\
Kimi K2.6 & 119/2263/247 & 53/675/68 & 2/0/0 \\
Doubao Seed 2.0 Code & 365/2207/57 & 132/651/13 & 1/1/0 \\
Doubao Seed 2.0 Lite & 294/2234/101 & 111/660/25 & 0/1/1 \\
Doubao Seed 2.0 Pro & 130/2497/2 & 79/715/2 & 0/2/0 \\
GPT-5.1 & 229/2296/104 & 112/655/29 & 1/1/0 \\
Gemini 2.5 Pro & 81/2270/278 & 46/668/82 & 1/1/0 \\
Doubao Seed 2.0 Mini & 105/2296/228 & 44/695/57 & 0/2/0 \\
\bottomrule
\end{tabular*}
\caption{\textbf{Decisions on negative test cases.} Grouped counts on negative test cases for all 19 configurations using the Pi harness. Each cell is $P/F/U$: valid \texttt{PASS}, valid \texttt{FAIL}, and no valid decision. The cell denominator is their sum, $N=P+F+U$. Ordinary errors combine base, partial, near-miss, and regression test cases.}
\label{vb:negative-group-counts}
\end{table*}

\begin{figure*}[!t]
\centering
\includegraphics[width=\textwidth]{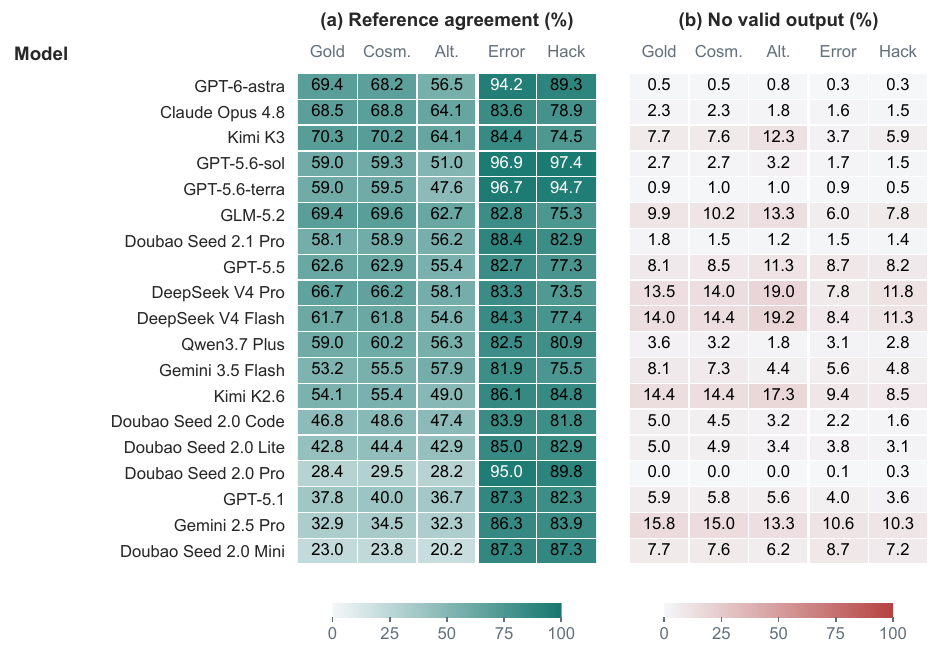}
\caption{\textbf{Test-case-level agreement and missing or invalid outputs.}
Rows follow the order of the configurations using the Pi harness in Table~\ref{vb:main-results}. Agreement is $P/N$ for the reference solution (Gold), cosmetic variants (Cosm.), and alternative implementations (Alt.), and $F/N$ for ordinary errors (Error) and hacks (Hack); missing or invalid output is $U/N$.
All 19 models are included. Each model uses the same group denominators from Tables~\ref{vb:group-counts} and~\ref{vb:negative-group-counts}.
The two process-violation test cases are omitted here and retained in Table~\ref{vb:negative-group-counts}.
Both color scales span 0--100\%. These test-case-weighted diagnostics differ from the task-averaged leaderboard metrics.}
\label{fig:vb-candidate-diagnostics}
\end{figure*}

The heatmap in Figure~\ref{fig:vb-candidate-diagnostics} separates reference agreement from missing or invalid outputs in each category.
This distinction prevents a low rate of accepted negative test cases from being interpreted automatically as successful rejection.
For positive test cases, Table~\ref{vb:positive-diagnostics} instead separates explicit false rejection from the absence of a valid decision.

For every model, explicit \texttt{FAIL} accounts for most unaccepted positive test cases, with rejection shares from 57.1\% to 100\%.
Valid-output rates across all groups range from 88.2\% to 99.9\%.
These output categories locate the observed disagreement but do not identify its cause; the analyses of task-level outcomes and verifier code in Section~\ref{sec:analysis} provide that additional evidence.
\begin{table*}[!tp]
\centering
\fontsize{10}{12}\selectfont
\setlength{\tabcolsep}{4pt}
\renewcommand{\arraystretch}{1.10}

\begin{tabular*}{\textwidth}{@{\extracolsep{\fill}}lrrrrrrr@{}}
\toprule
Model & $N$ & $N^+$ & $P$ & $F$ & $U$ & \shortstack{Rejection\\share (\%)} & \shortstack{Valid\\outputs (\%)} \\
\midrule
GPT-6-astra & 4993 & 1566 & 1012 & 545 & 9 & 98.4 & 99.6 \\
Claude Opus 4.8 & 4993 & 1566 & 1053 & 480 & 33 & 93.6 & 98.3 \\
Kimi K3 & 4993 & 1566 & 1069 & 354 & 143 & 71.2 & 94.2 \\
GPT-5.6-sol & 4993 & 1566 & 886 & 635 & 45 & 93.4 & 97.9 \\
GPT-5.6-terra & 4993 & 1566 & 871 & 680 & 15 & 97.8 & 99.1 \\
GLM-5.2 & 4993 & 1566 & 1055 & 336 & 175 & 65.8 & 92.1 \\
Doubao Seed 2.1 Pro & 4993 & 1566 & 907 & 636 & 23 & 96.5 & 98.5 \\
GPT-5.5 & 4993 & 1566 & 946 & 474 & 146 & 76.5 & 91.2 \\
DeepSeek V4 Pro & 4993 & 1566 & 997 & 325 & 244 & 57.1 & 89.1 \\
DeepSeek V4 Flash & 4993 & 1566 & 931 & 386 & 249 & 60.8 & 88.8 \\
Qwen3.7 Plus & 4993 & 1566 & 921 & 601 & 44 & 93.2 & 97.1 \\
Gemini 3.5 Flash & 4993 & 1566 & 876 & 589 & 101 & 85.4 & 94.3 \\
Kimi K2.6 & 4993 & 1566 & 832 & 494 & 240 & 67.3 & 88.9 \\
Doubao Seed 2.0 Code & 4993 & 1566 & 751 & 750 & 65 & 92.0 & 97.3 \\
Doubao Seed 2.0 Lite & 4993 & 1566 & 684 & 813 & 69 & 92.2 & 96.1 \\
Doubao Seed 2.0 Pro & 4993 & 1566 & 453 & 1113 & 0 & 100.0 & 99.9 \\
GPT-5.1 & 4993 & 1566 & 605 & 871 & 90 & 90.6 & 95.5 \\
Gemini 2.5 Pro & 4993 & 1566 & 526 & 812 & 228 & 78.1 & 88.2 \\
Doubao Seed 2.0 Mini & 4993 & 1566 & 353 & 1101 & 112 & 90.8 & 92.0 \\
\bottomrule
\end{tabular*}
\caption{\textbf{False rejection and valid outputs.} Outcomes on positive test cases and valid-output rates for all 19 configurations using the Pi harness. $N=4{,}993$ is the test-case total across all six groups; $N^+=1{,}566$ is the positive-test denominator for every model. Rejection share is $100F/(F+U)$, the fraction of unaccepted positives receiving an explicit \texttt{FAIL}. The valid-output rate includes all six groups and pools their test cases.}
\label{vb:positive-diagnostics}
\end{table*}

\section{Task-Level Error Analysis}
\label{app:traces}

\paragraph{Analysis population.}
The task-level analysis uses the archived records described in Section~\ref{sec:setup}, an October 3 export of selected judgments for GPT-6-astra, Kimi K3, GPT-5.6-sol, Qwen3.7 Plus, GPT-5.5, GPT-5.6-terra, GPT-5.1, Claude Opus 4.8, and Gemini 2.5 Pro with the Pi harness.
Each configuration covers all 222 tasks and has 4,993 evaluation records, one per test case.
For these nine models, the archived aggregate scores and every $N/P/F/U$ count in the six test-case groups agree with the final release.
The archive predates the final rejudging, so this agreement establishes consistency at the aggregate level rather than identity of every task-level decision.
Gemini 3.5 Flash is excluded from this analysis because its archived judgments differ from the final aggregates; all 19 models remain in the leaderboard and grouped diagnostics.
The nine-model concentration statistics are computed from these archived task-level records; the four-condition analysis below uses the Astra trace batch.

\begin{figure*}[!tp]
\centering
\begin{minipage}[t]{0.48\textwidth}
\vspace{0pt}
\centering
\includegraphics[width=\linewidth]{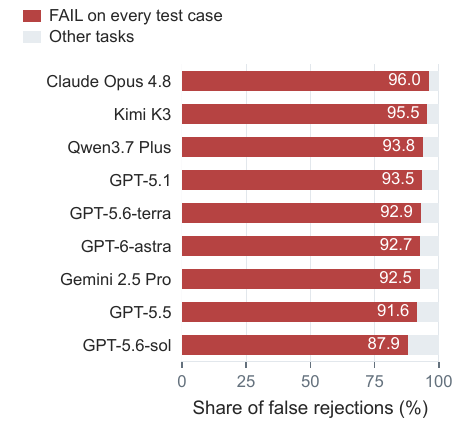}
\caption{\textbf{Concentration of false rejection.} For each of the nine Pi-harness configurations, the red segment gives the percentage of all positive \texttt{FAIL} decisions contributed by all-reject verifiers, which return \texttt{FAIL} on \emph{every} test case of their task. Across models this share stays between 87.9\% and 96.0\%. Gray segments represent the remaining false rejections. Appendix~\ref{app:traces} specifies the archived analysis records.}
\label{vb:concentration}

\end{minipage}\hfill
\begin{minipage}[t]{0.48\textwidth}
\vspace{0pt}
\makeatletter\def\@captype{table}\makeatother
\begin{minipage}{\linewidth}
\centering
\fontsize{10}{12}\selectfont
\setlength{\tabcolsep}{2.5pt}
\begin{tabular*}{\linewidth}{@{\extracolsep{\fill}}llrrr@{}}
\toprule
Harness & Effort & Tasks & Pos.\ \texttt{FAIL} & Share \\
\midrule
Pi harness & LOW & 75 & 551/587 & 93.9 \\
Pi harness & HIGH & 70 & 529/554 & 95.5 \\
Claude Code & LOW & 68 & 515/552 & 93.3 \\
Claude Code & HIGH & 70 & 535/564 & 94.9 \\
\bottomrule
\end{tabular*}
\caption{\textbf{False-rejection concentration in the Astra trace batch.} Tasks counts all-reject verifiers, which return \texttt{FAIL} on every test case of their task, out of 222. Pos.\ \texttt{FAIL} reports the positive \texttt{FAIL} decisions from these verifiers divided by all positive rejections in the condition; Share expresses this ratio as a percentage.}
\label{vb:final-concentration}
\end{minipage}

\par\vspace{12pt}
\begin{minipage}{\linewidth}
\centering
\fontsize{10}{12}\selectfont
\setlength{\tabcolsep}{3pt}
\begin{tabular*}{\linewidth}{@{\extracolsep{\fill}}llrrr@{}}
\toprule
Harness & Effort & None & Some & All \\
\midrule
Pi harness & LOW & 76 & 23 & 123 \\
Pi harness & HIGH & 71 & 16 & 135 \\
Claude Code & LOW & 69 & 22 & 131 \\
Claude Code & HIGH & 71 & 19 & 132 \\
\bottomrule
\end{tabular*}
\caption{\textbf{Positive acceptance by task.} Counts out of 222 tasks per Astra condition. None, Some, and All refer only to positive test cases. Each condition has one zero-acceptance task whose verifier nevertheless accepts a negative case, so None differs from the task count in Table~\ref{vb:final-concentration}.}
\label{vb:acceptance-distribution}
\end{minipage}

\end{minipage}
\end{figure*}

\paragraph{Concentration definition.}
For model $m$, let $\mathcal A_m$ contain the tasks whose submitted verifier is all-reject, that is, returns a valid \texttt{FAIL} on every test case for that task.
Tasks with any missing or invalid decision are excluded from $\mathcal A_m$.
If $F^+_{m,t}$ is the number of positive test cases rejected on task $t$, the test-case-weighted contribution is
\begin{equation}
 Q_m=\frac{\sum_{t\in\mathcal A_m}F^+_{m,t}}
 {\sum_{t\in\mathcal D}F^+_{m,t}}.
\end{equation}
Figure~\ref{vb:concentration} and Table~\ref{vb:final-concentration} report $100Q_m$.
This observed response pattern on the test suite does not establish that the verifier returns a constant decision on all possible inputs.

An equal-task-weight version replaces each $F^+_{m,t}$ with $F^+_{m,t}/|\mathcal C_t^+|$ in the numerator and denominator.
Its range across the nine models is also 87.9--96.0\%; for GPT-6-astra it is 91.1\%, compared with the test-case-weighted 92.7\%.
Thus, concentration is not explained solely by some tasks having more positive test cases.

\paragraph{Concentration in the Astra trace batch.}
We apply the same definition to the four conditions of the Astra trace batch.
These comprise 888 selected verifiers and 19,972 judgments, with 222 tasks and 4,993 judgments per condition.
Every judgment returns a valid binary decision.
Table~\ref{vb:final-concentration} shows that verifiers returning \texttt{FAIL} on every test case for a task account for 93.3--95.5\% of all false rejections.
These results come directly from the Astra trace batch, independently of the archived records above.
The four conditions concern one model and do not estimate variation across independent generation seeds.

The equal-task-weight concentration is 93.9\%, 95.1\%, 93.4\%, and 94.4\% for Pi harness/LOW, Pi harness/HIGH, Claude Code/LOW, and Claude Code/HIGH, respectively.
Table~\ref{vb:acceptance-distribution} separately reports whether each verifier accepts none, some, or all positive cases.

\paragraph{Task-source variation.}
Appendix~\ref{app:task-types} reports the task-type rankings for all 19 Pi harness models.
Within state tasks, GPT-6-astra's positive acceptance varies widely across source benchmarks.
The large within-type variation cautions against attributing all state-task failures to one mechanism.
Within the archived nine-model records, removing any single source still leaves overall TPR below TNR, while four models have the opposite relation within tool-call tasks.
The global asymmetry and the source-dependent differences can therefore coexist.

\section{Code and Generation-Trace Cases}
\label{app:cases}

We first compare three models on the same test cases to distinguish errors in input handling, checking rules, and execution.
We then use the Astra trace batch to examine source-file restrictions, tool-call handling, and incomplete testing.
Supplementary DNA and booking cases use archived verifier versions, while the AppWorld case concerns compatibility with the code forms of positive test cases.
The cases are selected to explain particular observed failures, not to estimate mechanism prevalence.
Hidden decisions are used only for post-generation analysis.

\subsection{Comparing Models on the Same Test Cases}
\label{app:cross-model-cases}

Table~\ref{vb:cross-model-cases} compares GPT-5.6-sol, Claude Opus 4.8, and Kimi K3, all using the Pi harness, on one task of each type.
Within each task, the models are evaluated on identical test-case contents.
The comparisons use the archived records described in Section~\ref{sec:setup}; they illustrate different failure mechanisms rather than explain the task-type rankings as a whole.

\begin{table*}[!tp]
\centering
\fontsize{10}{12}\selectfont
\setlength{\tabcolsep}{5pt}
\renewcommand{\arraystretch}{1.08}
\begin{tabular*}{\textwidth}{@{\extracolsep{\fill}}lrrr@{}}
\toprule
Model & Positive \texttt{PASS} & Negative \texttt{FAIL} & No valid decision \\
\midrule
\multicolumn{4}{l}{\textbf{Tool-call: update a user profile (ACE preference 17)}} \\
GPT-5.6-sol & 0/6 & 18/18 & 0 \\
Claude Opus 4.8 & 6/6 & 6/18 & 0 \\
Kimi K3 & 6/6 & 16/18 & 0 \\
\midrule
\multicolumn{4}{l}{\textbf{State: change an order address ($\tau^2$ retail/34)}} \\
GPT-5.6-sol & 0/9 & 8/8 & 0 \\
Claude Opus 4.8 & 9/9 & 8/8 & 0 \\
Kimi K3 & 9/9 & 8/8 & 0 \\
\midrule
\multicolumn{4}{l}{\textbf{Code: parse an HTTP date (Django-11848)}} \\
GPT-5.6-sol & 4/4 & 11/11 & 0 \\
Claude Opus 4.8 & 4/4 & 11/11 & 0 \\
Kimi K3 & 3/4 & 6/11 & 6 \\
\bottomrule
\end{tabular*}
\caption{\textbf{Model comparisons on identical test cases.} All three models use the Pi harness. Fractions count correct decisions over all cases of the corresponding class. The six missing decisions for Kimi K3 on Django-11848 are timeouts on one positive and five negatives, which remain in the denominators.}
\label{vb:cross-model-cases}
\end{table*}

\paragraph{Input compatibility and omitted checks produce different errors.}
On ACE preference 17, GPT-5.6-sol attempts to parse a supplied profile string as structured data and rejects all six positives when parsing fails.
Claude Opus 4.8 accepts all six positives but checks the requested location change without adequately checking preservation of the name and email, accepting 12 of 18 negatives.
Kimi K3 checks these preserved fields and accepts only two negatives; both contain the same repeated-call payload, recorded in two construction categories.
The comparison separates failure to interpret the supplied input from failure to enforce all required conditions.

\paragraph{Early rejection can prevent state verification.}
On retail/34, GPT-5.6-sol requires a tool result to parse as structured data before reaching its handling of a bare user-ID string.
The positive cases provide such a string, so all nine receive \texttt{FAIL} before any environment replay.
Claude Opus 4.8 and Kimi K3 handle this input and correctly judge all nine positives and eight negatives.
This contrast identifies an input-processing failure, rather than an incorrect judgment of the resulting order state.

\paragraph{Timeouts differ from incorrect decisions.}
On Django-11848, GPT-5.6-sol and Claude Opus 4.8 correctly judge all 15 test cases.
Kimi K3 correctly judges nine, while one positive and five negatives time out.
The records establish incomplete execution, but do not establish the cause of these timeouts.
Together, the cases show why score differences require separate examination of input handling, task-specific checks, and completion of execution.

\subsection{Attributing Rejections to Unjustified Checks}
\label{app:rule-cases}

We inspect the public task requirements, generated checks, positive test cases, recorded judgments, and generation traces together.
Table~\ref{vb:rule-defects} summarizes cases from the Astra trace batch for which the public requirements support attributing rejection to an unjustified check.
Its lower block contains the verifiers whose positive self-tests never returned a complete \texttt{PASS} during generation; its upper block contains the rest.
This is a purposive review of failures, not a random sample or an estimate of mechanism prevalence.
The counts refer to task--configuration verifier instances and their recorded decisions; repeated judgments across configurations and surface variants are not independent error mechanisms.
Source inspection identifies sufficient rejection conditions, without intervention experiments establishing that each condition is the unique cause of failure.

\subsection{Source-File Restrictions and Self-Test Coverage}
\label{app:django-case}

\paragraph{Django-11734.}
The Pi harness/LOW verifier for \nolinkurl{vgb-t-d8bd9183d1119aa2} rejects patches modifying any file named \texttt{\_\_init\_\_.py}.
The task concerns incorrect model references in \texttt{OuterRef}; its patch specification does not prohibit modifying the normal source module \nolinkurl{django/db/models/fields/__init__.py}.
All four positive test cases modify this module and receive \texttt{FAIL} before environment execution.
These cases comprise a reference repair and surface variants, not four independent repair strategies.

The agent's positive self-test instead modifies \nolinkurl{django/db/models/sql/query.py}.
After fixing subprocess compatibility, patch formatting, and environment reset issues, the trace records positive \texttt{PASS} and negative \texttt{FAIL} outcomes.
Those repairs make the chosen test executable but do not challenge the file restriction.
The relevant events are at zero-based native trace records 16, 26, 28, and 29, with the successful output at record 29.
The other three Astra configurations accept all four positives and reject all ten negatives on this task; this is an existing configuration comparison, not an intervention on the faulty verifier.

\paragraph{Matplotlib-14623.}
For \nolinkurl{vgb-t-25932ca825bd7ab6}, both Claude Code verifiers reject legitimate modifications to \nolinkurl{lib/mpl_toolkits/mplot3d/axes3d.py}.
The LOW verifier's allowed directories omit \nolinkurl{lib/mpl_toolkits/}; the HIGH verifier treats Python files under \texttt{lib} but outside \nolinkurl{lib/matplotlib/} as protected.
Each rejects four positive cases, yielding eight decisions on four distinct cases.
LOW rejects locally without environment execution; HIGH invokes the environment and rejects at the protected-file check before the relevant functional test.
Successful generation-time tests modify allowed locations such as \nolinkurl{axes/_base.py} or \texttt{ticker.py}.
Changing the repair within those locations does not test whether the directory restriction itself is justified.

\subsection{Tool-Call Restrictions}

The Pi harness/LOW verifiers for \nolinkurl{banking_knowledge/task_037}, \nolinkurl{task_040}, and \nolinkurl{task_041} reject authorized user-side tool calls allowed by the public tool protocol.
They respectively require the assistant role, explicitly reject the user-call wrapper, or omit its handling branch.
Their released task identifiers end in \texttt{12aafca53ad113b0}, \texttt{2de811fb047dcfa6}, and \texttt{36e8a311f52b2bb9}.
Each rejects nine positives, for 27 decisions.
For task 041, the verifier permits authorization through \nolinkurl{give_discoverable_user_tool}, but omits \nolinkurl{call_discoverable_user_tool}; the positive cases use that wrapper to query card information.
The intended-positive self-tests use assistant calls and do not exercise this allowed variation.

The Claude Code/LOW verifier for task 044, shown in Figure~\ref{vb:rule-case}, and Pi harness/LOW verifier for task 045 omit the permitted read-only tool \nolinkurl{get_user_dispute_history_7291}.
Their checks reject legitimate auxiliary reads as unsupported operations, producing 18 positive \texttt{FAIL} decisions.
Their self-tests omit the same read.
These five verifiers attempted positive self-tests, but none obtained a complete positive \texttt{PASS}; they therefore do not exemplify successful self-testing.

\subsection{Incomplete Category Vocabulary}
\label{app:retail-case}

The Pi harness/HIGH verifier for \texttt{retail/33} and both HIGH verifiers for \texttt{retail/34} implement the office-item category with vocabularies that omit \texttt{Headphones}.
Their positive cases contain this item, causing the selected order to fail the generated category/count condition.
Across these three verifiers, 27 positive \texttt{FAIL} decisions correspond to 18 distinct cases.
The two tasks share an order background, so they are not independent semantic replications.

For \texttt{retail/34}, the hidden positive cases contain Mechanical Keyboard, Headphones, and Office Chair as the three office items.
The Claude Code/HIGH verifier recognizes only Mechanical Keyboard and Office Chair, so its count falls below the required three items.
All nine positive cases receive \texttt{FAIL} without public environment execution.
The omitted category member is sufficient for rejection, although some cases may also encounter an earlier check.

Across these three retail and five banking verifiers, all 72 positive judgments have zero environment executions.
All eight generation traces acknowledge incomplete positive validation.
These eight verifiers form the lower block of Table~\ref{vb:rule-defects}.

\paragraph{Other permitted representations.}
The remaining cases in Table~\ref{vb:rule-defects} concern an unjustified ISO-only currency restriction, numeric-only device identifiers, and incomplete decoding of TeX spacing.
Their tests use the representations already accepted by the generated checks, leaving the omitted alternatives untested.
As with the code cases, a positive outcome observed during generation does not establish that the final source version completed full validation: the source version at every self-test was not reconstructed.

\paragraph{Compatibility-qualified case.}
The four Astra verifiers for Django-12155 require the old internal helper \texttt{trim\_docstring} to remain available, although the inspected repair replaces its handling with \texttt{inspect.cleandoc}.
This check accounts for 16 positive rejections on four distinct cases, but interpreting it as unnecessary depends on compatibility requirements not fully specified in the available materials.
We therefore exclude it from the 12-task, 16-verifier attribution count in Table~\ref{vb:rule-defects}.

\subsection{Incorrect Handling of a Specified Default}

The Qwen3.7 Plus/Pi harness case in Appendix Figure~\ref{fig:demo-dna} uses task \nolinkurl{vgb-t-d78e768308367722}, a DNA-analysis task from BFCL.
The actual generation prompt states that \texttt{mutation\_type} is optional and defaults to \texttt{insertion}; only the two sequence arguments are required.
The generated checker compares expected argument values with values obtained using
\begin{quote}
\small\ttfamily actual\_args.get(key, "")
\end{quote}
Consequently, an omitted optional parameter becomes the empty string rather than its defined default.
The positive test case \nolinkurl{vgb-c-e6ede484b58a7c02} omits this parameter in an insertion call and receives \texttt{FAIL}, while the designated fully specified solution receives \texttt{PASS}.

The same verifier searches for the presence of six expected calls and rejects extra calls only when their tool name is unexpected.
Negative test case \nolinkurl{vgb-c-9c2f334fedb91aae} preserves those calls and adds another call using the same tool; the generated verifier returns \texttt{PASS}.
The default omission is explicitly justified by the tool schema.
The treatment of extra calls is assessed relative to the reference acceptance rule and should not be generalized into an assumption that every redundant operation is invalid.

The visible self-tests use fully specified calls, an incomplete subset, and a raw-text representation of the fully specified calls.
They do not test omission of the default argument.
The case therefore links a specific comparison error to a behavior specified in the tool definitions but absent from the self-tests.

\subsection{Self-Tests That Share a Parsing Assumption}

The GPT-5.6-sol/Claude Code case uses task \nolinkurl{vgb-t-0480046e6a2a021f}, an ACEBench booking-cancellation task.
Its recorded generation includes a test described as a function call, constructed by placing a JSON dictionary inside the function's parentheses.
The test passes.
The reference solution in the test suite instead expresses the arguments as Python keywords, including \texttt{bookingId='R123'}.

The generated parser extracts the text inside the parentheses and attempts to decode it with \texttt{json.loads} or \texttt{ast.literal\_eval}.
A keyword-argument list is neither a JSON value nor a Python literal expression, so neither operation parses that text.
The verifier maps this failure to \texttt{FAIL}; the designated reference solution receives a recorded rejection with confidence 0.99.
The self-test passes because it uses precisely the representation accepted by the parser.

The verifier input format provides raw and parsed call fields, but the generation prompt gives no concrete grammar for the raw field.
This case supports input incompatibility and inadequate self-test coverage, rather than a claim that the model ignored an explicit syntax requirement.
It comes from the Claude Code configuration; Appendix Figure~\ref{fig:demo-sol} shows a related raw-input parsing failure under Codex.

\subsection{Compatibility with Code Forms of Positive Test Cases}
\label{app:appworld-case}

The GPT-6-astra/Pi harness LOW case uses AppWorld task \texttt{29caf6f\_1}, released as \nolinkurl{vgb-t-08167d1b4c9984dd}.
The task requires sending a movie recommendation.
The candidate solution in the selected positive test case, \nolinkurl{vgb-c-2d2f17e345dbfd3b}, imports \texttt{ApiCollection} from \nolinkurl{appworld.collections.apis} and \texttt{Requester} from \nolinkurl{appworld.requester}, then defines a \texttt{solution} function.

The verifier's \texttt{safe} function walks the candidate's Python AST.
Its check at source line 9 returns false for any \texttt{ast.Import} or \texttt{ast.ImportFrom} node, together with \texttt{ast.ClassDef}, \texttt{ast.With}, and \texttt{ast.AsyncWith}.
The caller returns a negative decision when \texttt{safe(code)} is false (line 23), before reaching \texttt{vgb\_env.run\_public} (line 28).
This positive test case therefore encounters a rejection condition before the verifier executes the proposed solution or examines its resulting state.
This identifies a specific incompatibility with a code form used by a positive test case; it does not establish that every AppWorld failure has this cause.
The archived prompt does not fully specify the solution-code entry convention, so this case is not counted among the unjustified checks in Table~\ref{vb:rule-defects}.

The generation trace also records a successful repair that leaves this restriction untested.
The model constructs three self-tests: a correct recommendation, an incorrect recommendation, and an empty input.
Their initial decisions are \texttt{PASS}, \texttt{PASS}, and \texttt{FAIL}.
The model identifies persistent state between invocations, inserts a reset before execution, and obtains \texttt{PASS}, \texttt{FAIL}, and \texttt{FAIL} on the repeated tests.
The positive self-test uses direct API calls without either import, so the test passes without exercising the condition that rejects the reference solution in the test suite.

Native trace records 21--22 contain the initial tests and results, record 23 adds the reset, and records 25--26 contain the repeated tests and results.

\subsection{Comparative Generation Trajectories}
\label{app:demo-trajectories}

The comparisons below expand the three findings in Section~\ref{sec:analysis}.
Figure~\ref{fig:demo-django} contrasts an overbroad file restriction with checks that preserve legal source changes.
Figure~\ref{fig:demo-dna} compares successful self-tests with a valid variation they omit.
Figure~\ref{fig:demo-retail} shows how an incomplete category vocabulary rejects valid solutions by miscounting the items specified by the task.
Figures~\ref{fig:demo-sol} and~\ref{fig:demo-opus} then illustrate the native-harness differences in Table~\ref{vb:native-harness}: input restrictions and access to a usable expected state.

\begingroup
\definecolor{vbDemoInk}{HTML}{263D52}
\definecolor{vbDemoBlue}{HTML}{496D8A}
\definecolor{vbDemoTeal}{HTML}{397970}
\definecolor{vbDemoRust}{HTML}{A65F49}
\newsavebox{\vbDemoBox}
\newenvironment{vbDemoPanel}[2]{%
  \par\addvspace{6pt}\def\vbDemoAccent{#1}%
  \setlength{\fboxsep}{7pt}\setlength{\fboxrule}{0.4pt}%
  \begin{lrbox}{\vbDemoBox}%
  \begin{minipage}{\dimexpr\linewidth-2\fboxsep-2\fboxrule\relax}%
  \fontsize{10}{12}\selectfont\raggedright\color{vbDemoInk}%
  \setlength{\parindent}{0pt}\setlength{\parskip}{4pt}%
  {\color{#1}\bfseries #2}\par\smallskip
}{%
  \end{minipage}\end{lrbox}%
  \noindent\fcolorbox{\vbDemoAccent!40}{\vbDemoAccent!5}{\usebox{\vbDemoBox}}\par
}
\newcommand{\vbDemoTitle}[2]{%
  \setlength{\fboxsep}{8pt}%
  \noindent\colorbox{vbDemoInk}{%
  \begin{minipage}{\dimexpr\linewidth-2\fboxsep\relax}%
  \color{white}\raggedright
  {\fontsize{11}{13}\selectfont\bfseries #1}\par\vspace{4pt}%
  {\fontsize{10}{12}\selectfont #2}%
  \end{minipage}}\par
}
\newcommand{\vbDemoCode}[1]{\par\smallskip{\fontsize{9}{11}\selectfont\ttfamily\raggedright #1\par}\smallskip}
\newcommand{\vbDemoPass}{\textcolor{vbDemoTeal}{\texttt{PASS}}}
\newcommand{\vbDemoFail}{\textcolor{vbDemoRust}{\texttt{FAIL}}}

\clearpage
\begin{figure*}[p]
\centering
\begin{minipage}{\textwidth}
\vbDemoTitle{Case 1. Protecting tests while allowing legal source changes}{GPT-6-astra\quad\textbar\quad Pi harness and Claude Code, LOW\quad\textbar\quad Django-11734}
\begin{vbDemoPanel}{vbDemoBlue}{Task requirement}
Fix incorrect model references in queries using \texttt{OuterRef} with \texttt{exclude()} or negation.
The candidate is a source-code patch; normal Django source files may be modified.
The comparison asks whether protecting tests also requires excluding legal source changes.
\end{vbDemoPanel}
\par\addvspace{6pt}
\begin{minipage}[t]{0.487\linewidth}\vspace{0pt}
\begin{vbDemoPanel}{vbDemoBlue}{Pi harness: restrict file names}
The agent adds a rejection condition containing:
\vbDemoCode{path.endswith((\\
\hspace*{1em}'conftest.py','sitecustomize.py',\\
\hspace*{1em}'\_\_init\_\_.py'))}
This also excludes ordinary source modules named \texttt{\_\_init\_\_.py}, before their behavior can be checked.
\end{vbDemoPanel}
\end{minipage}\hfill
\begin{minipage}[t]{0.487\linewidth}\vspace{0pt}
\begin{vbDemoPanel}{vbDemoBlue}{Claude Code: protect tests and check behavior}
The agent protects the test infrastructure while allowing ordinary Django source changes.
It checks that test files remain unchanged, applies source changes, and runs both a task-specific \texttt{OuterRef} check and existing query and expression tests.
It does not ban all files named \texttt{\_\_init\_\_.py}.
\end{vbDemoPanel}
\end{minipage}
\par\addvspace{6pt}
\begin{minipage}[t]{0.487\linewidth}\vspace{0pt}
\begin{vbDemoPanel}{vbDemoTeal}{Self-testing: execution repairs succeed}
The positive self-test modifies \texttt{query.py}.
The agent fixes subprocess compatibility, malformed patch headers, and missing environment reset.
Subsequent complete-verifier tests return:
\begin{tabular*}{\linewidth}{@{\extracolsep{\fill}}lr@{}}
Chosen positive patch & \vbDemoPass\\
Incorrect patch & \vbDemoFail\\
Empty patch & \vbDemoFail
\end{tabular*}
The successful patch does not exercise the file-name restriction.
\end{vbDemoPanel}
\end{minipage}\hfill
\begin{minipage}[t]{0.487\linewidth}\vspace{0pt}
\begin{vbDemoPanel}{vbDemoTeal}{Self-testing: functional checks succeed}
The agent observes successful execution of its behavior check and the existing regression tests.
Its complete-verifier self-tests return:
\begin{tabular*}{\linewidth}{@{\extracolsep{\fill}}lr@{}}
Two intended positive patches & \vbDemoPass\\
Incorrect patch & \vbDemoFail\\
Empty or malformed patch & \vbDemoFail
\end{tabular*}
Both positive self-tests also modify \texttt{query.py}; broader repair-location coverage does not explain this contrast.
\end{vbDemoPanel}
\end{minipage}

\begin{vbDemoPanel}{vbDemoRust}{Hidden evaluation: the same legal source changes}
All four positive test cases modify \nolinkurl{django/db/models/fields/__init__.py}.
The Pi verifier rejects them before environment execution; the Claude Code verifier accepts them and still rejects every negative case.
\smallskip
\begin{tabular*}{\linewidth}{@{\extracolsep{\fill}}lrr@{}}
Generated verifier & Positive cases accepted & Negative cases rejected\\
\hline
Pi harness & 0/4 & 10/10\\
Claude Code & 4/4 & 10/10
\end{tabular*}
\end{vbDemoPanel}
\begin{vbDemoPanel}{vbDemoBlue}{What the comparison shows}
Both agents obtain successful self-tests at the same repair location, but only one verifier excludes another legal location.
The contrast connects the findings in Sections~\ref{sec:rules} and~\ref{sec:selftests}: an unnecessary restriction can reject several correct solutions while remaining unchallenged by successful self-tests.
\end{vbDemoPanel}
\end{minipage}
\caption{\textbf{Accepting legal source changes need not weaken rejection of tested violations.} The four positives comprise a reference repair and surface variants. The contrast concerns the generated checks, not a demonstrated improvement in self-test coverage.}
\label{fig:demo-django}
\par\vspace{5pt}
\begin{minipage}{\linewidth}\fontsize{10}{12}\selectfont
These selected cases explain observed failures and successes; they do not estimate how often each mechanism occurs or isolate the effect of a harness component.
Each comparison connects the task requirements, generated checks, self-tests, and hidden judgments.
Paired harness comparisons use the same model, task, and test cases.
Self-tests were chosen during generation; hidden test cases were evaluated afterward.
Code excerpts omit unrelated checks, and prose summarizes the observed generation steps.
\end{minipage}
\end{figure*}
\clearpage
\begin{figure*}[p]
\centering
\begin{minipage}{\textwidth}
\vbDemoTitle{Case 2. Explicit defaults versus permitted omission}{Qwen3.7 Plus\quad\textbar\quad Pi harness\quad\textbar\quad DNA analysis}
\begin{vbDemoPanel}{vbDemoBlue}{Task requirement}
The task requests insertion, deletion, and substitution analyses for two DNA sequences.
The tool definition makes \texttt{mutation\_type} optional, with default \texttt{insertion}; \texttt{sequence} and \texttt{reference\_sequence} are required.
An insertion call therefore has the same meaning with the default supplied or omitted.
\end{vbDemoPanel}
\begin{vbDemoPanel}{vbDemoBlue}{Generated check: missing arguments become empty strings}
\vbDemoCode{for key, val in expected\_args.items():\\
\hspace*{1em}actual\_val = actual\_args.get(key, "")}
The verifier compares this value with the expected argument.
For an omitted \texttt{mutation\_type}, it compares \texttt{""} with \texttt{"insertion"} instead of applying the specified default.
\end{vbDemoPanel}
\par\addvspace{6pt}
\begin{minipage}[t]{0.487\linewidth}\vspace{0pt}
\begin{vbDemoPanel}{vbDemoBlue}{Generation-time positive: explicit default}
Every call in the positive self-test supplies \texttt{mutation\_type}.
For an insertion call, the relevant argument is:
\vbDemoCode{"mutation\_type": "insertion"}
The comparison receives the expected value.
\end{vbDemoPanel}
\end{minipage}\hfill
\begin{minipage}[t]{0.487\linewidth}\vspace{0pt}
\begin{vbDemoPanel}{vbDemoBlue}{Hidden positive: omitted default}
A valid insertion call omits the optional argument:
\vbDemoCode{"analyze\_dna\_sequence": \{\\
\hspace*{1em}"reference\_sequence": "AGCTTAGCTA",\\
\hspace*{1em}"sequence": "AGCTTAGCTA"\\
\}}
The generated comparison instead receives an empty string.
\end{vbDemoPanel}
\end{minipage}
\par\addvspace{6pt}
\begin{minipage}[t]{0.487\linewidth}\vspace{0pt}
\begin{vbDemoPanel}{vbDemoTeal}{Self-tests: all expected outcomes observed}
\begin{tabular*}{\linewidth}{@{\extracolsep{\fill}}lr@{}}
Fully specified calls & \vbDemoPass\\
Missing required calls & \vbDemoFail\\
Fully specified raw calls & \vbDemoPass
\end{tabular*}
The agent tests an incomplete solution and another input representation, but both positive tests explicitly supply the default.
\end{vbDemoPanel}
\end{minipage}\hfill
\begin{minipage}[t]{0.487\linewidth}\vspace{0pt}
\begin{vbDemoPanel}{vbDemoRust}{Hidden evaluation: legal omission rejected}
\begin{tabular*}{\linewidth}{@{\extracolsep{\fill}}lr@{}}
Positive cases accepted & 5/9\\
Positive cases rejected & 4/9
\end{tabular*}
The four rejections contain two distinct payloads, omitting the default in the first or second insertion call.
They repeatedly trigger the same comparison error.
\end{vbDemoPanel}
\end{minipage}

\begin{vbDemoPanel}{vbDemoBlue}{What the comparison shows}
The test variations the agent selected leave the incorrect requirement unchanged: every successful insertion example explicitly supplies its default.
The hidden positives vary exactly that permitted property.
This illustrates Section~\ref{sec:selftests}: tests must challenge whether a condition is required, not only demonstrate acceptance of examples that satisfy it.
\end{vbDemoPanel}
\end{minipage}
\caption{\textbf{Successful self-tests leave the specified default untested.} The Qwen3.7 Plus verifier passes its generation-time tests yet rejects a variation explicitly permitted by the tool definition.}
\label{fig:demo-dna}
\end{figure*}
\clearpage
\begin{figure*}[p]
\centering
\begin{minipage}{\textwidth}
\vbDemoTitle{Case 3. An incomplete category vocabulary rejects valid solutions}{GPT-6-astra\quad\textbar\quad Claude Code, HIGH\quad\textbar\quad retail/34}
\begin{vbDemoPanel}{vbDemoBlue}{Task requirement}
Remove three office items while retaining hiking items.
If partial cancellation is unavailable, change the order to the user's NYC address instead.
The generated verifier uses item names to identify the relevant order.
\end{vbDemoPanel}
\begin{vbDemoPanel}{vbDemoBlue}{Generated check: a keyword list determines the item category}
The verifier requires three office items.
Its category vocabulary includes \texttt{laptop}, \texttt{computer}, \texttt{monitor}, \texttt{keyboard}, \texttt{chair}, and \texttt{headset}, but omits \texttt{headphones}.
\end{vbDemoPanel}
\begin{vbDemoPanel}{vbDemoBlue}{Hidden positive cases: task category versus generated check}
\begin{tabular*}{\linewidth}{@{\extracolsep{\fill}}lcc@{}}
Item & Office item in the task & Matched by the verifier\\
\hline
Mechanical Keyboard & yes & yes\\
\textcolor{vbDemoRust}{\textbf{Headphones}} & yes & \textcolor{vbDemoRust}{\textbf{no}}\\
Office Chair & yes & yes\\
\hline
\textbf{Office-item count} & \textbf{3} & \textcolor{vbDemoRust}{\textbf{2}}
\end{tabular*}
Cycling Helmet and Hiking Boots are also present.
Because the keyword list omits Headphones, the generated check fails to recognize all three office items and rejects the order.
\end{vbDemoPanel}
\begin{vbDemoPanel}{vbDemoRust}{Recorded judgments}
All nine positive test cases receive \texttt{FAIL} before public environment execution.
The omitted category member is sufficient for static rejection, although some cases may also encounter an earlier check.
\end{vbDemoPanel}

\begin{vbDemoPanel}{vbDemoBlue}{What the comparison shows}
The generated vocabulary excludes an item that belongs to the task's office-item category.
The resulting count fails the verifier's own condition, making a valid order appear inconsistent with the task.
This illustrates the unjustified checks analyzed in Section~\ref{sec:rules}.
\end{vbDemoPanel}
\end{minipage}
\caption{\textbf{An omitted category member produces an incorrect item count.} The task identifies three office items, but the generated vocabulary recognizes only two. All nine positive test cases receive \texttt{FAIL} without public environment execution.}
\label{fig:demo-retail}
\end{figure*}
\clearpage
\begin{figure*}[p]
\centering
\begin{minipage}{\textwidth}
\vbDemoTitle{Case 4. Successful self-tests miss an incompatible parsing condition}{GPT-5.6-sol\quad\textbar\quad Codex and Pi harness\quad\textbar\quad Booking cancellation}
\begin{vbDemoPanel}{vbDemoBlue}{Common task and verifier input}
Cancel five reservations with the specified times and policies.
The input provides \texttt{parsed}, containing structured calls and arguments, and \texttt{raw}, containing the calls as emitted.
The supplied description does not require \texttt{raw} to be JSON.
\end{vbDemoPanel}
\par\addvspace{6pt}
\begin{minipage}[t]{0.487\linewidth}\vspace{0pt}
\begin{vbDemoPanel}{vbDemoBlue}{Codex: require an additional JSON parse}
After checking the structured calls, the verifier decodes the raw field using:
\vbDemoCode{raw\_value = json.loads(raw)}
It requires this result to satisfy the task checks and match the structured calls.
A raw string that is not JSON causes rejection even when the supplied structured calls are valid.
\end{vbDemoPanel}
\end{minipage}\hfill
\begin{minipage}[t]{0.487\linewidth}\vspace{0pt}
\begin{vbDemoPanel}{vbDemoBlue}{Pi harness: use the structured calls}
The verifier uses \texttt{parsed} when it can read that field, and falls back to \texttt{raw} otherwise.
It checks the requested bookings and cancellation arguments without making a separate raw JSON parse a prerequisite.
The raw representation therefore does not override valid structured calls.
\end{vbDemoPanel}
\end{minipage}
\par\addvspace{6pt}
\begin{minipage}[t]{0.487\linewidth}\vspace{0pt}
\begin{vbDemoPanel}{vbDemoTeal}{Codex self-tests: observed success}
The agent fixes an unavailable Python command and runs the complete verifier.
\begin{tabular*}{\linewidth}{@{\extracolsep{\fill}}lr@{}}
Correct cancellations, reordered & \vbDemoPass\\
Incorrect cancellations & \vbDemoFail\\
Malformed input & \vbDemoFail
\end{tabular*}
The positive supplies only \texttt{parsed}; it bypasses the additional raw-input check.
\end{vbDemoPanel}
\end{minipage}\hfill
\begin{minipage}[t]{0.487\linewidth}\vspace{0pt}
\begin{vbDemoPanel}{vbDemoRust}{Pi harness self-test: execution blocked}
The agent also constructs a positive containing only \texttt{parsed} and attempts a complete-verifier test.
The shell fails because \texttt{unshare} is unavailable; no successful verifier result is observed.
This generation trace does not demonstrate more complete self-testing than Codex.
\end{vbDemoPanel}
\end{minipage}

\begin{vbDemoPanel}{vbDemoBlue}{Same hidden cases: valid structured calls and Python-style raw calls}
The positive supplies the structured calls together with raw text beginning:
\vbDemoCode{[RestaurantReservation\_cancelBooking(bookingId='R123', ...), ...]}
The remaining arguments and calls are omitted here.
This text is not JSON, so it triggers the additional Codex rejection condition.
\smallskip
\begin{tabular*}{\linewidth}{@{\extracolsep{\fill}}lrr@{}}
Generated verifier & Positive cases accepted & Negative cases rejected\\
\hline
Codex & 0/7 & 20/20\\
Pi harness & 7/7 & 20/20
\end{tabular*}
\end{vbDemoPanel}
\begin{vbDemoPanel}{vbDemoBlue}{What the comparison shows}
The verifier with successful self-tests rejects every hidden positive; the other verifier accepts all positives while rejecting the same negatives.
The code identifies an additional parsing condition, and the successful self-tests never exercise it.
This connects the harness comparison to Sections~\ref{sec:rules} and~\ref{sec:selftests}: the difference concerns the generated checks, while neither run tests the raw-field assumption.
\end{vbDemoPanel}
\end{minipage}
\caption{\textbf{Successful self-tests bypass the input check that rejects hidden positives.} The difference is input compatibility, not adherence to an explicitly specified raw-call grammar.}
\label{fig:demo-sol}
\end{figure*}
\clearpage
\begin{figure*}[p]
\centering
\begin{minipage}{\textwidth}
\vbDemoTitle{Case 5. Constructing an expected state versus observing one}{Claude Opus 4.8\quad\textbar\quad Pi harness and Claude Code\quad\textbar\quad retail/49}
\begin{vbDemoPanel}{vbDemoBlue}{Common task}
Exchange a mistakenly ordered IPX7 earbud for an earbud with the requested water resistance, choosing the cheapest other earbud in the same delivered order.
Both verifiers check the exchange arguments and compare the candidate's resulting environment state with an expected state.
\end{vbDemoPanel}
\par\addvspace{6pt}
\begin{minipage}[t]{0.487\linewidth}\vspace{0pt}
\begin{vbDemoPanel}{vbDemoBlue}{Pi harness: construct a target exchange}
The generated verifier derives exchange parameters from the candidate's read results and constructs an expected exchange transcript.
It relies on replaying that transcript to obtain a usable target state for comparison.
There is no recorded successful execution establishing this target during generation.
\end{vbDemoPanel}
\end{minipage}\hfill
\begin{minipage}[t]{0.487\linewidth}\vspace{0pt}
\begin{vbDemoPanel}{vbDemoBlue}{Claude Code: execute the target exchange}
The agent performs the correct exchange in the public environment and obtains the resulting state.
The generated verifier includes both an expected transcript and the observed state as a fallback, then compares candidate execution with the target and checks the required exchange arguments.
\end{vbDemoPanel}
\end{minipage}
\par\addvspace{6pt}
\begin{minipage}[t]{0.487\linewidth}\vspace{0pt}
\begin{vbDemoPanel}{vbDemoRust}{Pi harness: no successful execution test}
Shell commands fail because \texttt{unshare} is unavailable.
The agent's final response describes its sanity cases as:
\par\smallskip
\emph{``reasoned, since local exec is blocked''}
\par\smallskip
These checks do not establish that the expected transcript can be replayed successfully.
\end{vbDemoPanel}
\end{minipage}\hfill
\begin{minipage}[t]{0.487\linewidth}\vspace{0pt}
\begin{vbDemoPanel}{vbDemoTeal}{Claude Code: test using the observed state}
After live execution, the agent creates a local substitute for the environment client using the observed state and runs the verifier on six examples.
The correct exchange passes; incorrect, empty, and malformed examples fail.
These tests check verifier logic with a simulated environment response, not six complete live executions.
\end{vbDemoPanel}
\end{minipage}
\par\addvspace{6pt}
\begin{minipage}[t]{0.487\linewidth}\vspace{0pt}
\begin{vbDemoPanel}{vbDemoRust}{Pi harness: target comparison unavailable}
For the designated hidden positive, the recorded decision and final source identify rejection without a usable expected-state comparison.
The precise internal replay error is not recorded.
All nine positive cases are rejected.
\end{vbDemoPanel}
\end{minipage}\hfill
\begin{minipage}[t]{0.487\linewidth}\vspace{0pt}
\begin{vbDemoPanel}{vbDemoTeal}{Claude Code: same positives accepted}
The verifier accepts all nine positives and rejects all eight negatives.
The successful decisions are consistent with the generated checks having access to a target state obtained from actual environment execution.
\end{vbDemoPanel}
\end{minipage}

\begin{vbDemoPanel}{vbDemoBlue}{Comparison on the same test cases}
\begin{tabular*}{\linewidth}{@{\extracolsep{\fill}}lrr@{}}
Generated verifier & Positive cases accepted & Negative cases rejected\\
\hline
Pi harness & 0/9 & 8/8\\
Claude Code & 9/9 & 8/8
\end{tabular*}
The contrast complements Section~\ref{sec:incomplete}: constructing a check does not establish that the evidence it requires can be obtained.
Here, live execution establishes the expected state used by the successful verifier; simulated self-tests and hidden evaluation then provide different evidence about its decisions.
\end{vbDemoPanel}
\end{minipage}
\caption{\textbf{Live execution establishes the target state used for verification.} The paired traces distinguish a constructed replay, an observed environment result, and tests using a simulated response. They do not isolate which harness component caused the difference.}
\label{fig:demo-opus}
\end{figure*}
\endgroup

\end{document}